\documentclass[twocolumn,prb,showpacs,floatfix]{revtex4} 
\usepackage{graphicx}
\usepackage{dcolumn}
\usepackage{bm}

\begin{document}

\title{
Where is the spectral weight in magnetic neutron scattering in
  the cuprates?}
\author{J. Lorenzana}
\affiliation{SMC-INFM, ISC-CNR, Dipartimento di Fisica,
Universit\`a di Roma La Sapienza, P. Aldo Moro 2, 00185 Roma, Italy}
\author{G. Seibold}
\affiliation{Institut f\"ur Physik, BTU Cottbus, PBox 101344,
         03013 Cottbus, Germany}
\author{R. Coldea}
\affiliation{Oxford Physics, Clarendon Laboratory, Oxford OX1 3PU, United
Kingdom}

\date{\today}
\begin{abstract}
We present estimates in the Hubbard and Heisenberg
models for the spectral weight in magnetic neutron scattering
experiments on the cuprates. With the aid of spin-wave theory and
the time dependent Gutzwiller approximation we discuss how the
spectral weight is distributed among the different channels and
between high and low energies. In addition to the well known total moment
sum rule we discuss sum rules for each component
of the dynamical structure factor tensor which are peculiar for spin
1/2 systems. The various factors
that reduce the spectral weight at the relevant energies are singled
out and analyzed like: shielding factors,
weight at electronic energies, multimagnon process etc.
Although about 10\% $\sim$ 15\% of the naively expected weight is detected
in experiments after consideration of these factors
the missing weight is within the experimental uncertainties.
A large fraction of the spectral weight is hard to detect with present
experimental conditions.
\end{abstract}
\pacs{74.25.Ha 
78.70.Nx 
71.10.Fd,
74.72.-h 
}
\maketitle
\section{Introduction}
\label{sec:int}
Magnetic neutron scattering (MNS) in high temperature superconducting
cuprates usually detects about $10\% \sim 15\%$ of the spectral weight
dictated by a naive application of sum rules.
For example the total weight in a wide range of energy and momentum
in a recent experiment\cite{tra04} in La$_{2-x}$Ba$_x$CuO$_4$,
with $x=0.125$, is $\sim 0.22\mu_B^2$ whereas in the
insulating phase the naive expectation from sum rules  is that one
should find $2\mu_B^2$.  It is usually argued that this value should
be corrected for the hole destruction of moments by a $1-x$ factor
which still leaves a large fraction of spectral weight undetected.

This rises various problems in the interpretation of MNS. For example
it has been argued that the average of the dynamical susceptibility
weighted by the Fourier transform of the
magnetic interaction can be used to estimate the energy involved
in magnetic pairing and its temperature
dependence.\cite{sca98,dem98,dai99,dai00,kee02} Clearly to obtain
an absolute estimate the spectral weight problem needs to be sorted
out first. Furthermore modeling the dynamical
structure factor probed by MNS becomes rather problematic since
sensible theoretical models do satisfy sum rules. Indeed any
theoretical claim of intensity agreement with the measured
dynamical structure factor in absolute units needs to explain
how the sum rule is satisfied or why it is violated.
This is even more stringent in
spin only models for which neither $1-x$ factors nor shielding
corrections apply.

The purpose of this work is to explain this apparent discrepancy.
We provide theoretical estimates of the various factors which
correct the sum rule and estimate what fraction of the spectral
weight is accessible to present day experimental conditions.
Theoretical estimates are performed in the antiferromagnetic (AFM)
phase using the Heisenberg and the Hubbard model combining spin-wave theory,
numerical results and the time dependent Gutzwiller
approximation (TDGA)\cite{sei01,sei03,sei04b} and in the doped phase
in the Hubbard model within the TDGA.
Apart from the mentioned $1-x$ factor we discuss the so called
``shielding factors'' due to an incomplete formation of magnetic
moments. We estimate the spectral weight loss to
electronic transitions at energies too high to be detectable by present
day inelastic magnetic neutron scattering  experiments and also the weight in
multimagnon processes which is either at too high energies or is so
broad in energy and momentum that it is not detectable in unpolarized
neutron scattering experiments. After consideration of all these
factors we arrive to the conclusion that within the
experimental uncertainties the sum rule is not violated (which is
reassuring) on the other hand a major fraction of the spectral
weight is very hard to detect with present experimental conditions.

The outline of the paper is at follows. In Sec.~\ref{mns} we shortly
review the theory of magnetic neutron scattering and the relevant sum
rules to fix notations. This section has also a pedagogical character.
Apart from the well known total moment sum rule we discuss
sum rules for each component of the dynamical structure factor tensor
which, to the best of our knowledge, have not been applied in the
present context. We also highlight some simple experimental facts that are
usually assumed as granted in experimental works, like domain averages
(Sec.~\ref{da}), but often overlooked in theoretical works.
In Sec.~\ref{hf} we discuss the spectral weight distribution
in the undoped case and in Sec.~\ref{ahf}  we discuss the doped case.
We conclude in Sec.~\ref{con}.

\section{Magnetic neutron Scattering}
\label{mns}
We start with a short review of magnetic neutron scattering to fix
notations and discuss the sum rules that are relevant to our problem.

The magnetic neutron scattering cross section is given by:\cite{lov84}
\begin{eqnarray}
  \label{eq:sigma}
\frac{d^2\sigma}{d\Omega dE'}&=&N\frac{k'}k \left(\frac{\gamma
    r_e}{2\mu_B}\right)^2 \left| F({\bm q})\right|^2 e^{-2W({\bm q})}\nonumber  \\
&\times& \sum_{\alpha\beta}
    (\delta_{\alpha\beta}-\hat q_\alpha\hat q_\beta)
    S^{\alpha\beta}(\bm{q},\omega )
\end{eqnarray}
where
  $e^{-2W({\bm q})}$ is the Debye-Waller factor, ${\bm
  q}\equiv {\bm k}-{\bm k}'$ ($\hat {\bm q}\equiv {\bf q}/|{\bm q}|$),
${\bm k}$ (${\bm k}'$) is the initial
  (final) wave vector of neutrons and
%
 $(\gamma r_e/2\mu_B)^2=72.65\times 10^{-3}$barn$/\mu_B^2$.
The magnetic form factor is given by
$F({\bm q})=\int dr e^{i{\bm q}.{\bm r}} |\phi({\bm r})|^2$
 where $\phi({\bm r})$ is a Wannier orbital
 and we defined the dynamical structure factor tensor,
\begin{eqnarray}
  && S^{\alpha\beta}(\bm{q},\omega )=\frac{(g \mu_{B})^2}{NZ}\label{eq:sw}\\
&&\times\sum_{\mu\nu}  e^{-\beta E_{\mu}} \langle \mu |
S^{\alpha}_{-{\bm q}}|\nu\rangle \langle \nu |  S^{\beta}_{\bm q}
|\mu\rangle \delta(\hbar\omega -E_\nu+E_\mu),\nonumber
\end{eqnarray}
where $g$ is the Land{\'e} $g$-factor. For free electrons
$g=2.0023$. In a solid a different value may be appropriate
which may also depend on direction. For example Ref.~\onlinecite{wal90}
quotes $g=2.08$ in the plane and $g=2.36$ perpendicular to the plane
for a typical cuprate.  For simplicity we take an isotropic $g$
unless otherwise specified.
$S^{\alpha}_{\bm q}$ is the Fourier transform of the $\alpha$
component of the spin operator and
$Z$ is the partition function. The dynamical structure factor
$S^{\alpha\beta}(\bm{q},\omega)$ obeys detailed balance
$S^{\alpha\beta}(\bm{q},\omega)= e^{\beta\hbar\omega
}S^{\alpha\beta}(-\bm{q},-\omega)$.

\subsection{Sum rules}
We will discuss sum rules for effective models of the magnetic
dynamics. Because effective models restrict the Hilbert space
sum rules turn out to be model dependent and therefore should be
applied with care on modeling a real system, as discussed below.
 Our considerations are based on
the two most popular models in this context, namely the Heisenberg
and the Hubbard model, respectively.

The so called total moment sum rule is usually formulated within
a  Heisenberg model with spin $S$ and reads:
\begin{equation}
  \label{eq:m0}
M_0\equiv\frac{1}{N} \sum_{\bm{q}\alpha}\int_{-\infty}^{\infty}
d(\hbar\omega) S^{\alpha\alpha} (\bm{q},\omega)=(g \mu_{B})^2
S(S+1)
\end{equation}
This applies to a system where magnetic ions have one or several
partially filled orbitals
(for example a rare earth ion with a partially filled $f$-shell)
and well formed magnetic moments i.e. when
double occupancy of a given orbital is negligibly.

Within this work we will restrict to systems where ions have
only one partially filled orbital and the system can be model with a
one-band Hubbard model:
\begin{equation}
\label{eq:hubbard}
H=-t\sum_{\langle ij \rangle \sigma}c_{i\sigma}^{\dagger}c_{j\sigma}
- t'\sum_{\langle\langle ij\rangle\rangle\sigma}c_{i\sigma}^{\dagger}
c_{j\sigma}
+ U\sum_{i}
n_{i\uparrow}n_{i\downarrow}.\nonumber
\end{equation}
Here $c_{i\sigma}^{\dagger}$ ($c_{i\sigma}$) destroys (creates) an electron
with spin $\sigma$ at site
$i$, and $n_{i\sigma}=c_{i\sigma}^{\dagger}c_{i\sigma}$. $U$ is the
on-site Hubbard repulsion and both  nearest ($\sim t$) and
next-nearest ($\sim t'$) neighbor hopping has been included.
Most of our considerations apply also to other models where ions have only
one partially filled orbital per atom, like the usual three-band Hubbard model
for cuprates with Cu $d$ and O $p$ orbitals.

We define for later use the spin autocorrelation function
and the zeroth moment of the diagonal  components of the dynamical
structure factor as:
\begin{eqnarray}
\label{eq:srw}
S^{\alpha\alpha}(\omega)&\equiv& \frac{1}{N} \sum_{\bm{q}}  S^{\alpha\alpha} (\bm{q},\omega)\\
  \label{eq:m0as}
M_0^{\alpha}&\equiv&\int_{-\infty}^{\infty}
d(\hbar\omega) S^{\alpha\alpha} (\omega)
\end{eqnarray}

When one has one orbital per site (i.e. a
spin 1/2 system) more stringent sum rules than Eq.~(\ref{eq:m0})
apply. Indeed {\em each
component} of the dynamical structure factor satisfies a separate sum rule.
 For example from Eqs.~(\ref{eq:sw}),(\ref{eq:srw}),(\ref{eq:m0as}) one finds:
\begin{eqnarray}
  \label{eq:m0z}
  M_0^{z}&=&\frac{(g \mu_{B})^2}{N^2Z}
\sum_{\mu{\bm q}}  e^{-\beta E_{\mu}} \langle \mu |
S^{z}_{-{\bm q}}  S^{z}_{\bm q}|\mu\rangle \nonumber\\
&=& \frac{(g \mu_{B})^2}4   (n-2D)
\end{eqnarray}
where  $S^{z}_{\bm q}=\sum_{i}
e^{-i{\bm q}.{\bm r_i}}S^z_i$ with $S^z_i=(n_{i\uparrow}-
n_{i\downarrow})/2$ and we used that $n_{i\sigma}^2=n_{i\sigma}$.
We also defined the thermal ($\langle ... \rangle$) and spacial averages of
the orbital occupancy:
$$
n\equiv \frac1{N} \sum_{i\sigma}  \langle  n_{i\sigma} \rangle,$$
and double occupancy:
$$
D\equiv \frac1{N} \sum_{i}\langle   n_{i\uparrow}n_{i\downarrow}\rangle.
$$
Analogous proofs hold for the other  components:
\begin{equation}
  \label{eq:m0a}
M_0^{\alpha}=\mu_{B}^2 (n-2D)
\end{equation}
where we took  $g=2$. (We will occasionally restore $g$ below when
convenient for clarity). Eq.~(\ref{eq:m0a})
also follows
from the fact that the quantization axis in Eq.~(\ref{eq:m0z}) is arbitrary.

Eq.~(\ref{eq:m0}) is valid in the Heisenberg model where $D=0$ and $S$
is arbitrary.
Eq.~(\ref{eq:m0a}) in contrast is valid for $S=1/2$ systems
but without restriction in $D$.

The factor $(n-2D)$ in Eq.~(\ref{eq:m0a}) is the probability to find
an atom singly occupied and reflects the fact that doubly occupied or
empty atoms do not
produce magnetic scattering of neutrons reducing the total scattering
cross section. This is some times called ``shielding factor''.

One can also prove that the total weight of the off-diagonal components of
$S^{\alpha\beta}$ adds to zero. First notice that only the symmetric
part of  $S^{\alpha\beta}$ contributes to Eq.~(\ref{eq:sigma}):
\begin{eqnarray*}
&&\sum_{\alpha\beta}
    (\delta_{\alpha\beta}-\hat q_\alpha\hat q_\beta)
    S^{\alpha\beta}(\bm{q},\omega )\\
&&=\frac12  \sum_{\alpha\beta}
    (\delta_{\alpha\beta}-\hat q_\alpha\hat q_\beta)
    [S^{\alpha\beta}(\bm{q},\omega )+S^{\beta\alpha}(\bm{q},\omega )].
\end{eqnarray*}
Using the Lehmann representation Eq.~(\ref{eq:sw}) one can show that:
\begin{eqnarray}
  \label{eq:m0ab}
M_0^{\alpha\beta}&\equiv&\frac{1}{2N} \sum_{q}\int_{-\infty}^{\infty}
d(\hbar\omega) [S^{\alpha\beta}(\bm{q},\omega)+ S^{\beta\alpha}(\bm{q},\omega)]\nonumber\\
&=&\delta_{\alpha\beta} \mu_{B}^2 (n-2D)
\end{eqnarray}
which gives  us a sum rule for each component of
$S^{\alpha\beta}(q,\omega)$.

We are not aware of references quoting
the sum rule Eq.~(\ref{eq:m0ab}) although the result is so simple that
we doubt it is original.

If $S^z_{tot}=\sum_i S^z_i$ is a good quantum number
$S^{\alpha\beta}(q,\omega )$ becomes diagonal with $S^{xx}(q,\omega )=S^{yy}(q,\omega ).$\cite{lov84}
In this case we can write the cross section as
\begin{eqnarray}
  \label{eq:sigmas}
\frac{d^2\sigma}{d\Omega dE'}&=&N\frac{k'}k \left(\frac{\gamma
    r_e}{2\mu_B}\right)^2 \left| F(\bm{q}) \right|^2 e^{-2W(\bm{q})} \\
&\times&\left[ (1-\hat q_z^2)S^{zz}(\bm{q},\omega )+(1+\hat q_z^2)S^{xx}(\bm{q},\omega )
\right].\nonumber
\end{eqnarray}
We will restrict to systems where this expression applies.

\subsection{Ordered states and Bragg scattering}
In the presence of long-range magnetic order the system
shows magnetic elastic scattering.
We will consider phases in which spin rotational invariance is
broken  with order along the $z$ axis (i.e. stripes, AFM,
etc.). We assume a magnetic unit cell with $N_a^M$ atoms
($N_a^M=2$ for the AFM state at half-filling) at positions
${\bm R}+ {\bm \delta_i}$,  with ${\bm R}$
the cell position and ${\bm \delta}_i$ the position of the atoms
within the cell ($i=1,...,N_a^M$). The vectors ${\bm R}$ form a
Bravais lattice.
For such a magnetic structure we have:
$$\langle S^z_{\bm q}  \rangle=N \sum_{\bm{Q}_M} \delta_{\bm{q}-\bm{Q}_M} m_{\bm{Q}_M}$$
where the sum is over the magnetic reciprocal basis vectors  and we
defined:
\begin{equation}
  \label{eq:mdq}
m_{\bm{Q}_M}\equiv \frac1{N_a^M} \sum_{i}^{N_a^M} e^{i \bm{Q}_M \cdot
  \bm{\delta}_i} m_i,
\end{equation}
and the local site-dependent magnetization $m_i=\langle
n_{i\uparrow}-n_{i\downarrow}\rangle/2$.

It is convenient to define the fluctuation operator:
 $$\delta S^z_q\equiv S^z_q-\langle S^z_q  \rangle .$$

With these definitions the longitudinal structure factor
can be put as:
\begin{eqnarray}
  \label{eq:szafh}
S^{zz}(\bm{q},\omega )&=&(g\mu_B)^2[ N \sum_{{\bm Q}_M} m_{{\bm Q}_M}^2
\delta(\bm{q}-\bm{Q}_M)\delta(\hbar\omega)\nonumber\\
&+& \sum_{\nu}
|\langle 0|\delta S_{\bm q} |\nu\rangle|^2
\delta(\hbar\omega-E_{\nu}+E_0)]
\end{eqnarray}
where for simplicity we set $T=0$ and $\delta ({\bm q})$ is
Kronecker's $\delta$ whereas
$\delta(\hbar\omega)$ is Dirac's $\delta$. The first term
in the brackets describes Bragg peaks. The weight of the peaks is given by the
square of the Fourier transform of the magnetization inside the
magnetic unit cell Eq.~(\ref{eq:mdq}). The second term describes
inelastic scattering.   The inelastic part (only!) of the
dynamical structure factor is related to the
dynamical susceptibility via the fluctuation
dissipation theorem.
\subsection{Domain average}
\label{da}
For the case discussed above of a diagonal structure factor tensor,
the cross section involves the factors $(1-\hat q_\alpha^2)$.
These are
polarization factors for scattering with unpolarized neutrons
and are rooted in the dipolar interaction between neutrons
and the electron spin.
In experimental works often the polarization factors   $(1-\hat q_\alpha^2)$
are included in the definition of the dynamical structure factor and
an average over the  orientation of domains,
$\langle ...\rangle_{dom}$,  is done:
 \begin{eqnarray}
 \label{eq:seff}
  S^{eff}(\bm{q},\omega)&=&
 \sum_{\alpha}
    \langle (1-\hat q_\alpha^2)\rangle_{dom}
    S^{\alpha\alpha}(\bm{q},\omega)
\end{eqnarray}
For a paramagnet all directions are equivalent,
$S^{\alpha\alpha}(\bm{q},\omega)$
does not depend of $\alpha$  and  $ \langle
(1-\hat q_\alpha^2)\rangle_{dom} =2/3$.
 Ordered systems will be
characterized by an order parameter that is a vector
like the staggered magnetization.  In general a real sample will
consist of domains with different orientations of the order
parameter. For a distribution of orientations that is completely
isotropic in spin space  $ \langle (1-\hat q_\alpha^2)\rangle_{dom} =2/3$.

Consider now scattering from a quasi two-dimensional (2d) system
with the $c$ direction defined perpendicular to the plane and the $a$ and $b$
directions in the plane.
In quasi 2d-systems the most common experimental
configuration is that the planes are perpendicular to the incident
neutron beam and, depending on the energy, the component of
${\bm q}$ perpendicular to the
plane, $q_c$ may be larger than the components in the plane.
In the extreme case that $\hat q_c>>\hat q_a,\hat q_b$ one can put
[c.f. Eq.~(\ref{eq:seff})]:
 \begin{eqnarray}
 \label{eq:seffy}
  S^{eff}(\bm{q},\omega)&=& S^{aa}(\bm{q},\omega)+S^{bb}(\bm{q},\omega)
\end{eqnarray}
This is valid regardless of the domain distribution. In order to
evaluate this expression it is convenient to use a domain-dependent
reference system in which the $z$ axis follows the ordered moment
of the domain.  In the case of an {\em isotropic} distribution of domains
one recovers $(1-\hat q_\alpha^2)\rangle_{dom} =2/3$. That is:
\begin{equation}
  \label{eq:siso}
S^{eff}(\bm{q},\omega)=\frac23 [S^{xx}(\bm{q},\omega)+S^{yy}(\bm{q},\omega)]+\frac23 S^{zz}(\bm{q},\omega)
\end{equation}
where we have grouped the transverse contribution.
 On the other hand, one usually deals with systems that have easy
 planes or easy axes. In cuprates for example
the ordered moment is usually in the Cu-O plane or close to that plane.
For a distribution of the ordered moment which is isotropic {\em within the
  plane} and in the above scattering geometry
 \begin{equation}
 \label{eq:seffqz}
  S^{eff}(\bm{q},\omega)=
\frac12[S^{xx}(\bm{q},\omega)+S^{yy}(\bm{q},\omega)]
+S^{zz}(\bm{q},\omega).
\end{equation}
The 1/2 factor is due to the fact that one transverse mode is
perpendicular to the plane and becomes silent in this
configuration.

For reasons that will be clear below real experiments
detect mainly the transverse structure factor.
Comparing Eqs.~(\ref{eq:siso}),~(\ref{eq:seffqz})
 we see that in order to accurately estimate spectral weights
from experiment precise information is needed on the domain
distribution since the transverse component appears with different weight.

The condition $\hat q_c>>\hat q_a,\hat q_b$ is
rather extreme. Instead, in the present scattering geometry,
 $\hat q_c/\hat q_a$,
 $\hat q_c/\hat q_b$ will gradually decrease as the energy
is increased. This will  make the factors weighting the transverse
and the longitudinal part to depend on energy. Due to the larger
sensibility to the transverse part in a real experiment
this will lead to an apparent loss of spectral weight as the energy increases.

 In an
ideal experiment where all components are detected with equal sensitivity
one will not see a loss of spectral weight but
rather a transfer of spectral weight from the longitudinal to the
transverse part as the energy increases. Indeed
$S^{eff}(\bm{q},\omega)$ satisfies the sum rule
\begin{equation}
  \label{eq:m0eff}
M_0^{eff}\equiv\frac{1}{N} \sum_{\bm{q}}\int_{-\infty}^{\infty}
d(\hbar\omega) S^{eff} (\bm{q},\omega)=  2 (n-2D) \mu_{B}^2
\end{equation}
independently of what the distribution of domains or the scattering
geometry is.

\subsection{Experimental Considerations}
\label{sec:exper}

On analyzing experimental data one should take into account that
unpolarized INS experiments magnetic and nonmagnetic scattering
can not be unambiguously identified. In the insulating phase this
problem can be reduced by fitting the data with a simple model
like spin-wave theory. Due to the lack of a simple theory in the doped
phase a similar procedure is not possible.   In this case the usual
experimental practice is to report as
``magnetic scattering'' only those features which satisfy certain
criteria like being reasonably sharp in momentum space and show a
``magnetic like'' temperature dependence and form factor dependence
across different Brillouin zones.
 All the rest is assumed to be background of
unknown origin. Below we analyze the distribution of weight in the
insulator to get a hint of the distribution of spectral weight in the
different channels and where to expect sharp and where broad features.

\section{Half-filling}
\label{hf}

\subsection{Spectral weights in the Heisenberg Model}
\label{sec:heis}
We start by neglecting the shielding factors and
estimate the spectral weights in the Heisenberg model
within spin-wave theory (SWT). We will show below how to apply these results
when shielding factors are important corrections.
Since we are interested in gross distributions of
spectral weight for simplicity we consider a Heisenberg model
with nearest neighbor interactions and neglect other terms like
four-ring exchange.

\subsubsection{Longitudinal part}
\label{sec:longitudinal-swt}
In order to write the longitudinal dynamical structure 
factor we introduce the following notations.  
For an AFM the $S^z_i$ operator can be written within SWT as:
$$
S^z_i=(S-{\cal N}_i)e^{-i \bm{Q}_{AFM}\cdot\bm{r}_i}
$$
where ${\cal N}_i$ is the number operator for Holstein-Primakoff bosons
and $\bm{Q}_{AFM}=(\pi,\pi)$ taking the lattice constant $a=1$.
 For the Fourier
transform $S^z_{\bm q}=\sum_i e^{-i \bm{q} \cdot \bm{r}_i} S^z_i$
we have:
$$S^z_{\bm q}=NS \delta(\bm{q}-\bm{Q}_{AFM})-{\cal N}_{\bm{q}-\bm{Q}_{AFM}}$$
It is convenient to write  ${\cal N}_{\bm{q}-\bm{Q}_{AFM}}\equiv
\langle {\cal N}_{\bm{q}-\bm{Q}_{AFM}} \rangle+\delta{\cal
N}_{\bm{q}-\bm{Q}_{AFM}} $ with
 $\langle{\cal N}_{\bm{q}-\bm{Q}_{AFM}}\rangle=
N \Delta S \delta(\bm{q}-\bm{Q}_{AFM})$ and define
 $m_{{\bm Q}_{AFM}}\equiv m\equiv S-\Delta S$ as the reduced sublattice
magnetization.  Here
$$\Delta S=\frac1N\sum_{\bm{q}}
\frac{1}{2}\left[\frac{1}{\sqrt{1-\gamma^2_{\bm q}}}-1 \right]$$ is
the reduction in the ordered spin moment due to zero-point quantum
fluctuations and $\gamma_{\bm{q}}=[\cos q_x +\cos q_y]/2$ comes from the
Fourier-transform of the exchanges in the square lattice.

The one-magnon  dispersion relation is given by:
\begin{equation}
  \label{eq:wdqswt}
\hbar\omega_{\bm q }=2JZ_C\sqrt{1-\gamma^2_{\bm{q}}}
\end{equation}
 where $J$ is  the superexchange constant and $Z_C$ is a quantum
renormalization of the one-magnon energy, near-constant over the
Brillouin zone in a first approximation,\cite{can92p10131,qmc} and
estimated as 1.18 to order $1/S^2$ in spin-wave theory\cite{can92p10131}
or by series expansions.\cite{sin89p9760}

\begin{figure}[htbp]
\centering
\includegraphics[width=8cm,clip=true,bbllx=101,bblly=457,
bburx=482,bbury=763,angle=0,clip=]{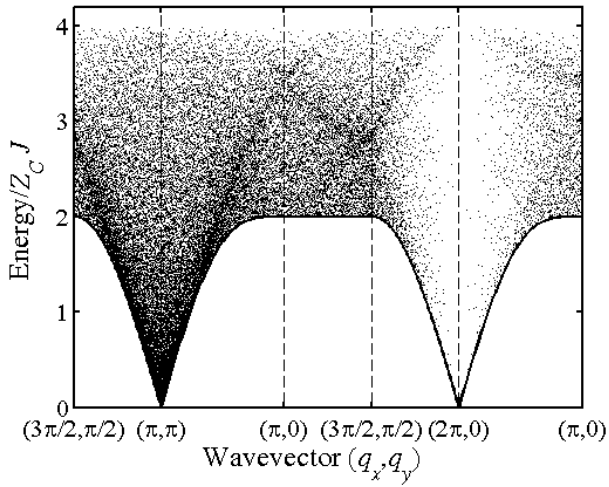}
\caption{Two-magnon scattering intensity [Eq.~(\ref{eq:2m})] as a
function of energy and wavevector along symmetry directions in the
Brillouin zone (solid lines in Fig.~\ref{fig:2m_sofq} inset).
Density of scattered points represents intensity. The lower
boundary traces the one-magnon dispersion relation
$\hbar\omega_{\bm q}$ and corresponds to events where one of the two
magnons has zero energy. The upper bound, $4Z_CJ$, is reached when
both magnons are on the antiferromagnetic zone boundary contour
(dashed square in Fig.~\ref{fig:2m_sofq} inset) where they have
maximum energy. \label{fig:2m_sqw}
}
\end{figure}

With these definitions we can write the
longitudinal dynamical structure factor as:
\begin{eqnarray}
  \label{eq:szaf}
S^{zz}_{Hei}(\bm{q},\omega )&=& g^2\mu^2_B[ N m^2\delta(\bm{q}-\bm{Q}_{AFM})\delta(\hbar\omega)\\
&+&\sum_{\nu} |\langle 0|\delta {\cal
N}_{\bm{q}-\bm{Q}_{AFM}} |\nu\rangle|^2
\delta(\hbar\omega-E_{\nu}+E_0)], \nonumber
\end{eqnarray}
where the first term in the brackets is the Bragg elastic contribution of the
N\'{e}el order and the second term is the inelastic contribution.

The sublattice magnetization is well known to be accurately given by
linear SWT,\cite{man91}  therefore we do not
expect significant changes on the elastic intensity if higher order
corrections are included.
In contrast how the inelastic part is distributed at low-energies
(a few $J$'s) is expected to be sensitive to such
corrections.

The inelastic part can be decomposed into a sum over inelastic processes
out of which the dominant term corresponds to two-magnon
scattering events with intensity given by\cite{hei81}
\begin{eqnarray}
&&S^{zz}_{2M}(\bm{q},\omega)= Z_{2M}\frac{g^2\mu^2_B}{2N}\label{eq:2m}\\
&&\times\sum_{\bm{q}_1,\bm{q}_2} f(\bm{q}_1,\bm{q}_2)
\delta(\hbar\omega-\hbar\omega_{\bm{q}_1}-\hbar\omega_{\bm{q}_2})
\delta(\bm{q}+\bm{q}_1-\bm{q}_2).\nonumber
\end{eqnarray}
The two-magnons have opposite spin $S_z=+1$ and $-1$ such that the
total spin $S^{z}_T$ is unchanged. The scattering cross-section
$f(\bm{q}_1,\bm{q}_2)$ depends on the wavevectors of the two
magnons via\cite{hei81}
\begin{equation}
f(\bm{q}_1,\bm{q}_2) = \sinh^2 \left( \theta_{\bm{q}_1}
-\theta_{\bm{q}_2}\right), ~~~~~
\theta_{\bm{q}}=\frac{1}{2}\tanh^{-1} \gamma_{\bm{q}},
\label{eq:f2m}
\end{equation}
We have included an {\it ad hoc} intensity renormalization factor $Z_{2M}$
to be discussed below.
  The lineshapes in Eq.~(\ref{eq:2m}) can be evaluated by direct
summation or Monte-Carlo methods and the result is illustrated by
the dotted areas in Fig.~\ref{fig:2m_sqw}. 

Two-magnon events
contribute a continuum band of scattering at energies above the
one magnon dispersion $\hbar\omega_{\bm q}$. In general the
intensity decreases with increasing energy and it is strongest for
wavevectors near the antiferromagnetic zone center $(\pi,\pi)$.

\begin{figure}[t]
  \centering
  \includegraphics[width=8cm,clip=true,bbllx=94,bblly=268,
  bburx=482,bbury=572,angle=0,clip=]{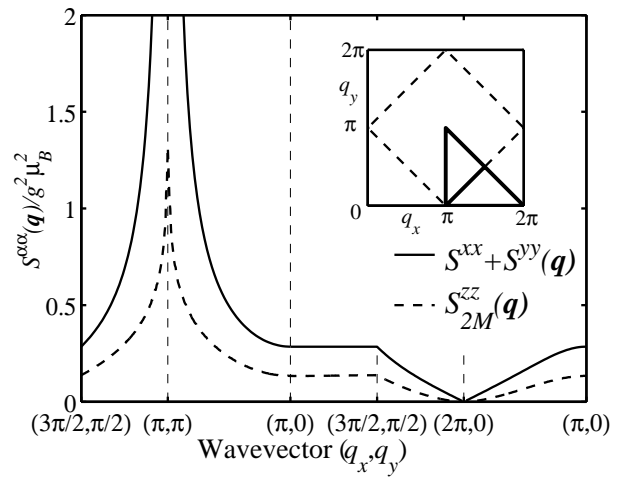} 
  \caption{Energy integrated spectral weight in the two-magnon
  continuum  (dashed line), $S^{zz}_{2M}(\bm{q})=\int
d(\hbar\omega) S^{zz}_{2M}(\bm{q},\omega)$, compared with the
weight in the one-magnon peak (solid line), from
Eq.~(\ref{eq:sxaf}). We have used $Z_{d}=0.57$ and $Z_{2M}=0.67$.
\label{fig:2m_sofq}}
\end{figure}

Fig.~\ref{fig:2m_sofq} shows how the energy-integrated intensity
varies in the Brillouin zone, generally following the same trend
as the one-magnon intensity. The two-magnon signal cancels at the
nuclear zone center $(2\pi,0)$ as expected from first-moment sum
rules for an isotropic antiferromagnet, also made apparent from
Eq.\ (\ref{eq:f2m}) where the cross-section for creating two
identical magnons cancels, i.e. for $\bm{q}_1=\bm{q}_2$ such that
$\bm{q}=0$, $f(\bm{q}_1,\bm{q}_1)=0$. 

\begin{figure}[bp]
  \centering
  \includegraphics[width=8cm,clip=true,bbllx=88,bblly=318,
  bburx=418,bbury=522,angle=0,clip=] {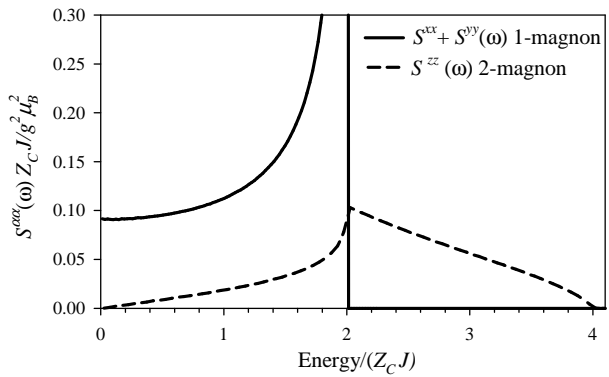}  
  \caption{Momentum integrated spectral function
of one- [Eq.~(\ref{eq:sxaf})] and two-magnon excitations
[Eq.~(\ref{eq:2m})]. We have used $Z_{d}=0.57$ and $Z_{2M}=0.67$.
Integrated areas are $(1+2\Delta S)Z_{d}S$ and $Z_{2M}\Delta
S(1+\Delta S)$, respectively.\label{fig:2mdos}}
\end{figure}

Fig.~\ref{fig:2mdos} shows
the wavevector-integrated intensity: it increases with increasing
energy, linearly at low energies, then reaches a maximum at the
one-magnon zone boundary and then decreases to zero at energies
above twice the maximum magnon energy. In fact most of the
two-magnon scatterig weight is at energies above the one-magnon
cutoff $2Z_CJ$. 

Due to the broadness of the distribution the
two-magnon spectral weight will be quite hard to detect in
unpolarized neutron experiments, although we note that recent
neutron experiments\cite{hub05} have observed such a
high-energy continuum of excitations in the square-lattice
spin-$5/2$ Heisenberg antiferromagnet Rb$_2$MnF$_4$ and the
observed intensities were in agreement with neutron scattering by
pairs of magnons as described by spin-wave theory. In cuprates, in
addition, a large fraction of the spectral weight is at energies
that are too high for neutron scattering.

Integrating Eq.~(\ref{eq:2m}) over energy and wavevector in a
Brillouin zone gives $Z_{2M}\Delta S(1+\Delta S)~g^2\mu^2_B$, whereas
the Bragg elastic scattering is $(S-\Delta S)^2~g^2\mu^2_B$. The
magnetization reduction can be
 evaluated numerically as $\Delta S = 0.197$. The corresponding value
 of $M_0^z$ is shown in Table~\ref{tab:sw1}.  Comparing with the Ising
 AFM, also shown in  Table~\ref{tab:sw1},
one sees that the effect of the transverse fluctuations is to reduce the
sublattice magnetization  with a concomitant  transfer of spectral
weight from the Bragg peak to the two-magnon continuum. Most of
the longitudinal scattering is in this spread-out two-magnon
continuum. The spin-only sum rule $M_0^z=1\mu_B^2$ is strongly
violated if intensity renormalization is neglected ($Z_{2M}=1$).
Notice that if transverse fluctuations are neglected (Ising limit) the
sum rule is exactly satisfied. We can anticipate a similar result in
the Hubbard model.

Including higher-order terms in spin-wave theory is
expected to produce: {\it i}) intensity-lowering of the
two-magnon response \cite{can93}
and {\it ii})  spread out at higher energies due to the contribution of
higher multimagnon process.\cite{san01}
For simplicity we assume the intensity renormalization  of the
two-magnon response to be momentum independent and
quantify it with the constant  $Z_{2M}$. 

Since the Bragg intensity is
given accurately by SWT one can get an upper bound for $Z_{2M}$ by
neglecting effect {\it ii}) and 
requiring
that the sum rule Eq.~(\ref{eq:m0z}) is fulfilled by Bragg
and two-magnon processes:
$Z_{2M}\sim 0.67$. Perturbative computations\cite{can93} suggest a
value not far from that suggesting that the weight in four magnon and
higher multimagnon processes
is not high, in agreement with numerical data.\cite{san01}
In the following to improve the values of the two-magnon intensity
we tentatively adopt $Z_{2M}= 0.67$ given the lack of better
estimates  (Table~\ref{tab:sw1}).

It is interesting to remark that the small weight at four magnon
and higher multimagnon processes is  in
striking contrast to the response relevant for
infrared experiments which show large weights
instead.\cite{per93,lor95,lor95b,kas98,gru00,lor99}
From a theoretical point of view, whereas magnon-magnon interactions
have dramatic consequences in the shape of the spectrum of
two-magnon excitation relevant for Raman and IR
data\cite{lor95,lor95b} the effect on the
two-magnon line shape relevant for magnetic neutron scattering appears
much modest.\cite{can93}
This different role of interactions can be traced back to the fact
that optical data probes
two magnons in different sublattices whereas magnetic neutron
scattering probes two magnons on the same sublattice and the
interaction is dominant in the first case.
The corresponding Green functions at the RPA level for both cases
are reported in Ref.~\onlinecite{lor95b}.
In optical data magnon-magnon interactions are responsible for
a shift of the main feature from $\sim 4J$ to
$\sim 3J$.\cite{can92p7127,lor95,lor95b} This effect was obtained in
Ref.~\onlinecite{lor95b} using a high-energy approximation.
One can easily see from the RPA equations for the equal-sublattice Green
function\cite{lor95b} that
at the same level of approximation the same effect does not show up in
the two-magnon neutron scattering line shape. We conclude that
interaction effects should be important for the intensity  renormalization but
they play a very different role here than the one played in optical data.

\begin{table}[b]
  \caption{\label{tab:sw1}
Sublattice magnetization $m_{{\bm Q}_{AFM}}$ and dynamical structure factor
spectral weights in units of $\mu_B^2$. We show exact values for the
Ising AFM and SWT values for the Heisenberg AFM in the longitudinal
channel ($z$) and in one transverse channel ($x$). SWT are the values
neglecting all renomalization factors ($Z=1$).
We also include the
 SWT intensity renormalized by the indicated $Z$
values. The transverse contribution is split in the contribution from
the magnetic zone and the nuclear zone as $M^x_{MBZ}+
M^x_{NBZ}$. Since we are considering the Heisenberg model ($D=0$) 
a ``1'' in the last column implies that the sum rule is exactly
satisfied within the model.
}
\begin{ruledtabular}
\begin{tabular}{cc|cccc}
     &$\alpha$&$m_{{\bm Q}_{AFM}}$&\text{Elastic}& \text{Inelastic}&$M_0^{\alpha}$\\
\hline
Ising& $z$    &    0.5     & 1      &  0                         &   1    \\
     & $x$    &            & 0      & 0.5+0.5                    &   1    \\
SWT  & $z$    &   0.3034   & 0.368  &0.941\footnotemark[1]       & 1.309  \\
     & $x$    &            & 0      & 1.068+0.325\footnotemark[2]& 1.393\footnotemark[2]  \\
$Z_{2M}=0.67$&$z$&0.3034   & 0.368  &0.632\footnotemark[1]       & 1  \\
$Z_d=0.57$\footnotemark[3]&$x$& & 0 & 0.609+0.185\footnotemark[2]& 0.794\footnotemark[2]
\end{tabular}
\end{ruledtabular}
\footnotetext[1]{Two-magnon contribution.}
\footnotetext[2]{One-magnon contribution.}
\footnotetext[3]{After Refs. \onlinecite{can92p10131,can93}}
\end{table}






\subsubsection{Transverse part}
\label{sec:swttran}

The transverse dynamical structure factor is dominated by
one-magnon scattering events with intensity given by\cite{man91}
\begin{eqnarray}
  \label{eq:sxaf}
S^{xx}_{Hei}(\bm{q},\omega )&=&g^2\mu^2_B~Z_{d}\frac{S}{2}
\frac{1-\gamma_{\bm q}} {\sqrt{1-\gamma^2_{\bm q}}}
\delta(\hbar\omega-\hbar\omega_{\bm q} ),
\end{eqnarray}
$Z_{d}$ is an
intensity-lowering renormalization factor of the one-magnon
cross-section due to zero-point fluctuations and magnon-magnon
interactions, both neglected at first order in spin-wave theory
($Z_{d}=Z_{\chi}Z_C$, where $Z_{\chi}$ is the renormalization of
the transverse magnetic susceptibility,
$\chi_{\perp}=Z_{\chi}(g\mu_B)^2/8J$).

The wavevector dependence of the one-magnon intensity is shown in
Fig.~\ref{fig:2m_sofq}. Although wave-vectors $(2\pi,0)\equiv(0,0)$ and
$(\pi,\pi)$ are related by symmetry the spectral weight goes to
zero at wave vector $(2\pi,0)$ and diverges at $(\pi,\pi)$. This
difference in intensity is reflected also in doped phases.

Integrating Eq.~(\ref{eq:sxaf}) over energy and wavevector in a
Brillouin zone gives the one-magnon intensity for one transverse
direction as $(1+2\Delta S)Z_{d}S g^2\mu^2_B/2$.
Experimental works often restrict to the magnetic Brillouin zone
around $(\pi,\pi)$ (dashed line in the inset of
Fig.~\ref{fig:2m_sofq}) hereafter ``the magnetic zone'' and neglect the
small weight on the magnetic Brillouin zone around $(0,0)$
hereafter ``the nuclear zone''. 
For comparison we split the one-magnon spectral weight in the two zones
$M_0^x= M_{MBZ}^x+ M_{NBZ}^x$.
 One finds the values reported in Table~\ref{tab:sw1}. 

In this case as
 in the longitudinal channel the sum rule is overestimated if one
 neglects the intensity renormalization.
 Estimates of the renormalization from  higher-order spin-wave theory
give\cite{can92p10131,can93} $Z_{d}=0.57$ at order $1/S^2$, or 0.61(4)
by series expansions.\cite{sin89p9760} Now the sum rule is
underestimated. The lacking weight is expected to lay in  three-magnon and
higher multimagnon processes.

An alternative way to estimate  $Z_{d}$ is to enforce
 Eq.~(\ref{eq:m0}) at large $S$ and extrapolate to $S=1/2$.
With the choice $Z_{d}=1-\Delta S/S=0.606$, the total
sum rule $S(S+1)$ [c.f. Eq.~(\ref{eq:m0})] is
exhausted by elastic $(S-\Delta S)^2$,
one-magnon $(1+2\Delta S)(S-\Delta S)$, and two-magnon scattering
$\Delta S (1+\Delta S)$ (without renormalization),
in units of $g^2\mu^2_B$. Such a
renormalization of the one-magnon intensity is very close to the
results quoted above. Strictly speaking, this argument applies only for $S>1/2$
since the sum rule for each channel Eq.~(\ref{eq:m0a}), which is
exclusive of spin 1/2 systems, is still strongly violated.
One the other hand the rapid convergence of the $1/S$ expansions
often produces accurate results for $S=1/2$ systems as seem to be the
case for $Z_d$.

\subsection{Shielding factor in the Hubbard Model}
\label{sec:1m2d}

We now turn to the more realistic Hubbard model.
In this section we estimate the shielding factor for $n=1$, within the 2d
single-band Hubbard model Eq.~(\ref{eq:hubbard}),
to gain some insight on its impact on the sum rule Eq.~(\ref{eq:m0ab}).
Here we take for simplicity
$t'=0$ and consider the effect of varying $U/t$. Parameters more
specific for the cuprates will be considered in Sec.~\ref{ahf}.
\begin{figure}[bp]
  \centering
  \includegraphics[width=8cm,clip=true]{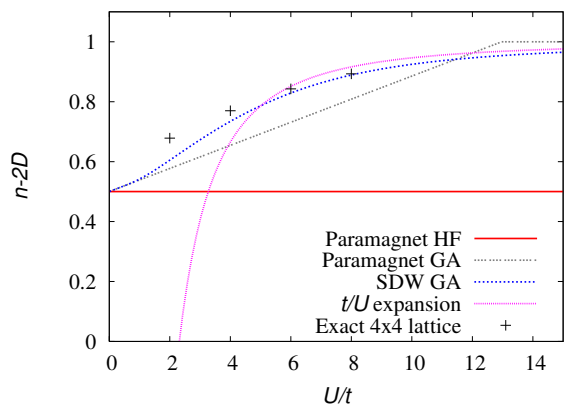}
  \caption{(Color online)
Single occupation probability (shielding factor)
for a half-filled Hubbard model in
  different approximations. The exact results in a $4\times 4$ site cluster are
  computed with the double occupancies given in
  Ref.~\protect\onlinecite{fan90}. The apparent lack of extrapolation
  of the exact results to the non-interacting limit at $U=0$ is
  due to a  finite size effect.}
  \label{fig:1m2d}
\end{figure}

The shielding factor is given by the single occupancy probability
$n-2D$.
For a non interacting system, i.e. $U=0$,  the double occupancy is just $(n/2)^2$
so one gets $n-2D=1/2$. Indeed half of the time an atom is singly
occupied (up or down)
and half of the time it is either empty or doubly occupied and hence does not
produce scattering.

For a paramagnetic state in the Hartree-Fock approximation (HF) one gets
the same result independently of $U/t$ (Fig.~\ref{fig:1m2d}) since
correlations are neglected.
 Correlations can be introduced by treating the paramagnet in the
 Gutzwiller approximation (GA)\cite{vol84} which
leads to a reduction in double occupancy as
 $$D=\frac{1}{4} \left(1- \frac{ U}{U_c}\right)$$
becoming zero at the Brinkman-Rice transition point $U_c/t=
128/\pi^2\simeq 12.97$.\cite{vol84,yos96}
The corresponding value of $n-2D$ is plotted also in
Fig.~\ref{fig:1m2d}. The  Brinkman-Rice point, however, is never reached
since for infinitesimal $U$ the paramagnetic state is unstable
towards a spin-density wave (SDW) which  can also be treated in the GA and gives the
value of  $n-2D$ shown in Fig.~\ref{fig:1m2d}.

Finally for large $U/t$ one can use a canonical transformation to
map the Hubbard model to a Heisenberg model.\cite{yos96,man91} It
is important to realize that the energies of the low-energy
excitations of both models, the ``physical'' Hubbard model and the
low-energy ``effective'' model coincide to leading order but the correlations
functions in general do not coincide.  In order to get
``physical'' correlation functions one needs to use the inverse
canonical transformation to transform back the ``effective''
ground state wave function to a physical wave function.
Indeed  within the Heisenberg model the double occupancy is
zero but this does not mean that the  double occupancy is zero in
the ``physical'' model. This is obviously very important on evaluating the
right hand side of Eq.~(\ref{eq:m0a}). Of course if we are
interested on evaluating the sum rule {\em within}  the Heisenberg
model it is legitimate to take $D=0$, as done above,
 but if we want to compare
with experiments (or with the ``physical'' model) one needs to
compute the ``physical'' double occupancy. Fortunately this is
very easy in the present case because we can use a trick to avoid
the back transformation. We use Hellman-Feynman theorem to write
the double occupancy in the Hubbard model as:
$$
D=\frac1N \frac{\partial E}{\partial U}
$$
with $E$ the ground state energy. For the
latter we use the fact that for large $U/t$ it coincides with  the
energy of the Heisenberg model  where very accurate numerical
estimates exist:
$$
\frac{E}N=-\alpha \frac{4t^2}U
$$
and the best estimate  for $\alpha$ is
$\alpha=0.6696$.\cite{man91} Here
we have substituted the superexchange constant by its definition
$J\equiv 4t^2/U$.
Since the Heisenberg model has only one parameter it is clear that
$\alpha$ does not depend on $J$ and we can perform the derivative to
obtain:
$$
D=\alpha \frac{4t^2}{U^2}.
$$
The corresponding value of the single occupancy probability $n-2D$ is
plotted in Fig.~\ref{fig:1m2d} and referred to as the ``$t/U$
expansion''. This value is asymptotically exact (within numerically
accuracy\cite{man91})  in the large $U/t$
limit and therefore takes into account all fluctuations effects.
Clearly the SDW treated within the GA  approximations gives a
fairly good approximation for $n-2D$ at large $U$ and interpolates
smoothly to the exact result at $U=0$ so we expect it to be quite
accurate in all the range of $U/t$ as can be seen also by
comparing with the exact results in a $4\times4$ cluster (after
Ref.~\onlinecite{fan90}).

We see that the sum rule for one diagonal component of the
dynamical structure factor Eq.~(\ref{eq:m0a}) changes smoothly
from $\mu_B^2/2$ in the non-interacting case to
 $\mu_B^2$ in the limit of $U=\infty$. It is reduced by $\sim 11\%$
(Table~\ref{tab:hub})
with respect to the full moment value for $U/t\sim 8$, as relevant for
cuprates.\cite{col01,sei05}

\subsection{Distribution of spectral weight in the Hubbard Model}
\label{sec:hub}
In the previous section we have evaluated the shielding factors that
appear due to the finiteness of $U$ and reduce the total spectral weight
compared to spin only models.
Another effect, which we study here, is that the total weight is split
into a low-energy part at energies of order $J$ and a high-energy part
at energies of order $U$. Present neutron scattering facilities can
measure the spectra up to energies of the order of a few tenth of
eV and therefore only the first part is detected. In addition we
discuss how the dynamical structure factor is modified for finite $U$
respect to the SWT result.

To estimate the  dynamical structure factor we use the time dependent
Gutzwiller approximation of
Refs.~\onlinecite{sei01,sei03,sei04b} [also called GA plus
random-phase-approximation (GA+RPA)] applied to a SDW state. We start
in the next section by showing  how measurements of the dispersion
relation in the insulator can be used to estimate $U$ and $t$.

\subsubsection{Transverse part}
\label{sec:hubtrans}

\begin{figure}[tbp]
  \centering
    \includegraphics[width=8cm,clip=true]{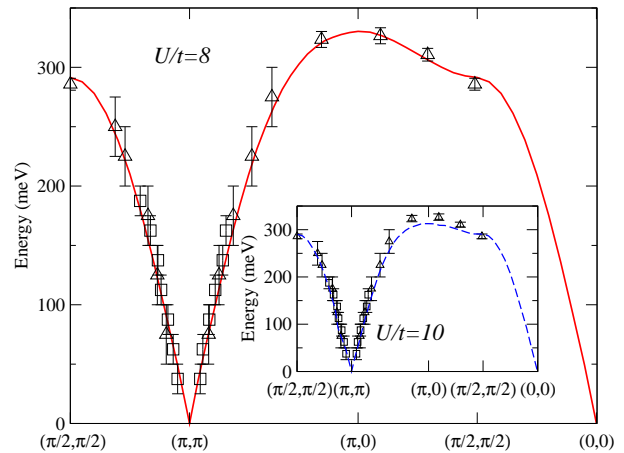}
  \caption{(Color online) Dispersion relation of the low-energy transverse
    excitations.  We show the experimental result for La$_2$CuO$_4$
after Ref.~\onlinecite{col01} and the  GA+RPA result
for $U/t=8$ and $t=335$meV. The inset shows that for
$U/t=10$ the fit is noticeably worse.
}
  \label{fig:edq}
\end{figure}

In the transverse channel MNS experiments reveal the spin-wave
excitations of the AFM.\cite{col01} Interestingly, whereas spin-wave
theory on the Heisenberg model Eq.~(\ref{eq:wdqswt})
predicts a flat dispersion
 between $(\pi,0)$ and $(\pi/2,\pi/2)$ a substantial dispersion has
 been measured (c.f. Fig.~\ref{fig:edq}). It has been argued that in
cuprates corrections to the Heisenberg model arising as higher orders
in a $t/U$ expansion are relevant.\cite{rog89,sch90,lem97,lor99,col01}
  The most important of such corrections is a term which cyclically
exchanges four spins on a plaquette. A sizable value for this term
has been revealed by analyzing phonon-assisted multimagnon infrared
absorption\cite{lor99} and the dispersion
relation\cite{col01} shown in  Fig.~\ref{fig:edq}. In particular
the dispersion between  $(\pi,0)$ and $(\pi/2,\pi/2)$ is mainly due
to this term. Since the dispersion has its origin in the finiteness of
$t/U$ it should show up in the transverse excitations of
the Hubbard model. The computation done in
GA+RPA is also shown in  Fig.~\ref{fig:edq}. One obtains a very
good fit of the dispersion and this provides an accurate way to
estimate the strength of the repulsion. We find $U/t=8$ in good
agreement with other estimates\cite{col01} whereas  $U/t=10$ gives
a too flat dispersion between  $(\pi,0)$ and $(\pi/2,\pi/2)$ (inset).
An equally good fit as the one show in the main panel can be achieved
with $U/t=8$, $t'/t=-0.2$ and $t=353.7$meV. The value of $t'/t$ plays an
important role in the doped phase\cite{sei04a,sei05}
and is close to a first principle estimate.\cite{pav01}

\begin{figure}[b]
  \centering
  \includegraphics[width=8cm,clip=true]{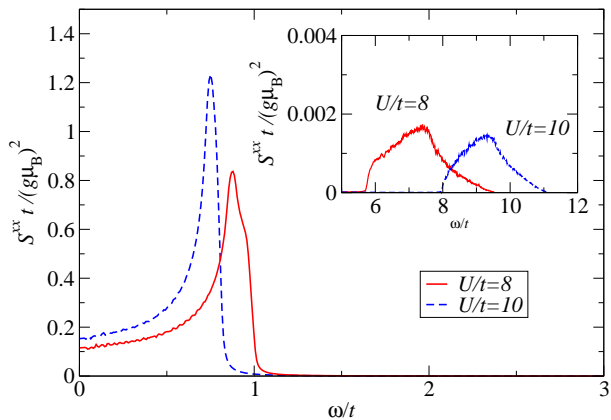}
  \caption{$S^{xx}(\omega)$ evaluated for the Hubbard model within
   GA+RPA for $U/t=8$ (solid line) and $U/t=10$ (dashed line) with
   $Z_d^U=1$ for an  $80\times80$ sites system.
   The inset shows the high energy contribution of $S^{xx}(\omega)$
   at energies $\omega \sim U$.
}
  \label{fig:swt}
\end{figure}

In Fig.~\ref{fig:swt} we show the momentum integrated spectral
function. We see that for large $U/t$ it approaches the SWT form
(Fig.~\ref{fig:2mdos}). For smaller
$U/t$ the maximum is not at the upper edge due to the modified
dispersion relation.
A small portion of the spectral weight  is at high-energy
due to spin-flip transitions from the lower to the upper Hubbard
band as shown in the inset. We will show that this effect is
much larger in the longitudinal channel.

Integrating the spectral intensity one gets the values of
$M_0^x$ shown in Table~\ref{tab:hub}.
GA+RPA interpolates between the extreme limits $U/t=0$
(where the sum rule is exactly obeyed) and $U \to \infty$ where
one recovers the results of linear spin-wave theory. As a consequence
the sum rule in the transverse channel is increasingly more violated
as $U/t$ increases with $M_0^x$ reaching values similar to the
unrenormalized SWT results (Table~\ref{tab:sw1}) for large $U$.

\begin{figure}[t]
  \centering
    \includegraphics[width=8cm,clip=true]{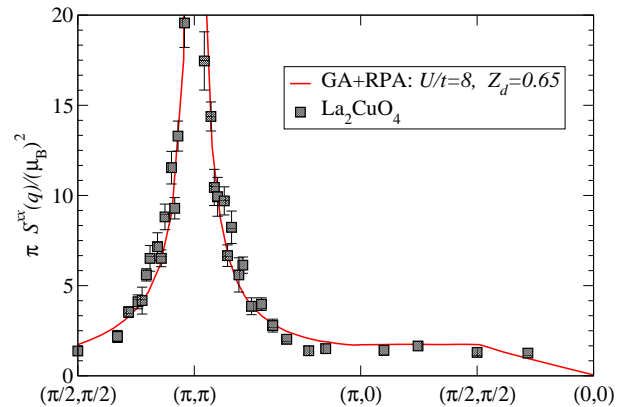}
  \caption{(Color online) Intensity in the spin-wave excitations
as a function of momentum for one transverse spin-channel. We show
the experimental result for La$_2$CuO$_4$ after
Ref.~\onlinecite{col01} and the  GA+RPA result for $U/t=8$ and $t=335$meV.
The GA+RPA intensity has been
renormalized by a factor $Z_{d}^{U=8t}=0.65$ and coincides
with the spin-wave result (shown in Fig. 3b of Ref.~\onlinecite{col01})
renormalized by $Z_{d}^{exp}=0.51$ in Eq.~(\ref{eq:sxaf}). }
  \label{fig:idq}
\end{figure}

In Fig.~\ref{fig:idq} we show the experimental intensities
as a function of wave vector together with the
 GA+RPA result. In order to fit the experimental results we
introduced an intensity renormalization $Z_d^{U=8t}$
analogous to the SWT  intensity renormalization factor.
With   $Z_d^{U=8t}=0.65$ the GA+RPA intensities are essentially
equivalent to the intensities reported in Fig. 3b
of Ref.~\onlinecite{col01} which have been
renormalized by $Z_{d}^{exp}=0.51$ {\em respect to LSWT}.

This increase in the value of  $Z_d^{U}$ reflects the fact
that cuprates are not in the strict  Heisenberg limit $U\rightarrow \infty$. 
In that limit the $Z_d$ for the RPA should coincide with the 
value of LSWT given the equivalence of the two approximations.
In the extreme case of a non interacting system 
RPA gives exact intensities and therefore $Z_d^{U=0}=1$. 
As $U$ is decreased from infinity the RPA becomes gradually more 
accurate and therefore $Z_d^{U}$ should increase respect to the LSWT
value as we indeed find. 

We caution that the fact that a system is not in the strict 
Heisenberg limit does not mean by itself that it can not be described 
by the Heisenberg model. Corrections to the intensity can be
introduced as explained in Sec.~\ref{sec:effect-shield-fact}.
We estimate this approach is accurate up to $U\sim 15 t$ where the 
dispersion is still Heisenberg like. For smaller $U$ corrections in
the dispersion relation due to four spin exchange and other non
Heisenberg processes become obvious (c.f. Fig.~\ref{fig:edq}).

\subsubsection{Longitudinal part}
\label{sec:hfhublon}
The longitudinal structure factor has a Bragg  part
and an inelastic part [c.f. Eq.~(\ref{eq:szafh})].
For the SDW only one magnetic reciprocal lattice vector contributes to
the sum in Eq.~(\ref{eq:szafh}) namely ${\bm Q}_{AFM}\equiv(\pi,\pi)$
(we take the lattice constant $a=1$).
  The Bragg weight is determined by the sublattice magnetization
 $m_{{\bm Q}_{AFM}}$. 

At lowest order RPA introduces longitudinal 
fluctuations in $S^{zz}$
but does not correct one-body expectation values like 
the sublattice magnetization. (On the contrary it introduces a 
correction in two-body expectation values \cite{sei01} like $D$). Therefore in the following we  
 consider the sublattice magnetization at the GA level together with the
effect of the  longitudinal fluctuations at high energies in $S^{zz}$. 
The sublattice magnetization can in principle be corrected by including
transverse fluctuations. We will discuss this below.

At the GA we obtain the values of the sublattice magnetization 
shown in Table~\ref{tab:hub}. These values lack the transverse
fluctuation corrections and hence converge to the classical value, 
0.5 in the large $U$ limit.

To help intuition it is useful to
distinguish between the permanent moment
$m=|\langle n_{i\uparrow}-n_{i\downarrow}\rangle|/2$,
which determines the Bragg weight, and an ``intrinsic''
moment which determines the shielding factor and which we define as
$m^*=\sqrt{\langle(n_{i\uparrow}-n_{i\downarrow})^2\rangle}/2=\sqrt{(n-2D)}/2$.
(Notice that with this definition a free fermion system has an
 intrinsic moment which we denote as ``trivial'').
The paramagnet with small $D$ mentioned in
Sec.~\ref{sec:1m2d} can be considered as a system with well
formed (non-trivial) intrinsic moments that are purely dynamical.
Indeed the shielding factor is close to 1 but there are no
Bragg peaks. In this regard the GA is more flexible than HF which is
not able to produce non-trivial intrinsic moments that are not
permanent.

Permanent moments are due to the breaking of spin-rotational symmetry.
The GA for the SDW smoothly interpolates between the itinerant
limit at small $U$ and the
localized limit at large $U$ with a permanent moment that increases and
reaches $1/2$ for $U\rightarrow\infty$. Contrary to the paramagnetic
case, quoted above, the difference between permanent and the intrinsic moment
in the GA tends to be very small at large $U$. Thus despite the ability
of the GA to distinguish the two kinds of moments this ability is not
effective at large $U$.  In this case
the permanent moment in the GA (c.f. Table~\ref{tab:hub}) is reduced with
respect to the fully polarized value mainly due to covalency effects.
On top of this the moment will be reduced due to 
transverse fluctuations  as discussed below.

\begin{table}[b]
  \caption{\label{tab:hub}
Spectral weights in units of $\mu_B^2$. We show exact values for the
Hubbard in the longitudinal channel ($z$) and in one transverse
channel ($x$). The inelastic weight is separated in the low-energy
part (Low) at energy $\sim J$ and the high-energy part  $M_{0,U}$
around energy $\sim U$.
We also show the sublattice magnetization $m_{{\bm Q}_{AFM}}$
and the shielding factor $n-2D$, both computed in the GA 
(without RPA correction).}
\begin{ruledtabular}
\begin{tabular}{dc|cccccc}
 U &$\alpha$& $m_{{\bm Q}_{AFM}}$ &\text{Elastic}&\multicolumn{2}{c}{Inelastic}&$M_0^{\alpha}$&$n-2D$\\
   &        &            &              &  Low       &$M_{0,U}^{\alpha}$&              &      \\
\hline
8  & $z$    &0.43        & 0.74         &  0         & 0.15           & 0.898        &0.889 \\
   & $x$    &            & 0            &1.1         &0.01          & 1.11          &      \\
10 & $z$    &0.456       & 0.83         &  0         & 0.096          & 0.930        & 0.923\\
   & $x$    &            &0             &1.196       &0.009          & 1.205          & \\
15 & $z$    &0.481       & 0.927        &  0         & 0.0397         & 0.966        & 0.965\\
   & $x$    &            &0             &1.278       &0.002          & 1.28           & \\
\end{tabular}
\end{ruledtabular}
\end{table}

\begin{figure}[t]
  \centering
  \includegraphics[width=8cm,clip=true]{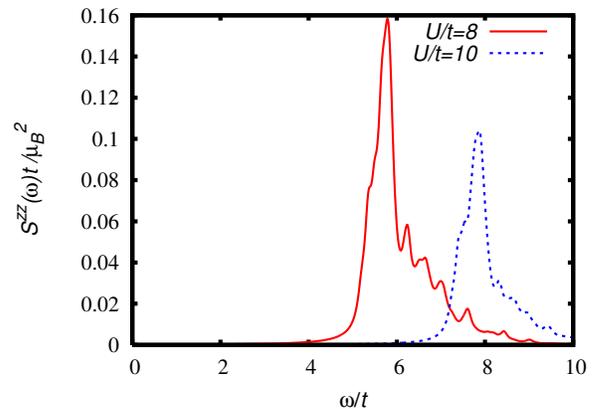}
  \caption{(Color online)  Inelastic part of
$S^{zz}(\omega)$ in the GA+RPA for different values of $U/t$ in
an  $8\times8$ sites system.
}
\label{fig:sw}
\end{figure}

The covalency reduced Bragg peak does not exhaust the sum rule as can
be seen by comparing the ``Elastic'' column with the last column in
Table~\ref{tab:hub}. Longitudinal fluctuations, captured by the GA+RPA
approach, produce weight at energies of order $U$ as shown
in Fig.~\ref{fig:sw}. This is not related to
two-magnon processes (as in Sec.~\ref{sec:longitudinal-swt})
but to fluctuations in the length of the intrinsic moment.
The weight in the high-energy part
$M_{0,U}$ is also reported in Table~\ref{tab:hub}.
Adding the Bragg and the $M_{0,U}^z$ contribution one obtains the
values of $M_0^z$ reported in Table~\ref{tab:hub}. Comparing with the
last column we see that at this level of approximation the sum rule is 
exhausted (actually slightly overshot) by the Bragg contribution
without transverse fluctuations reductions plus the contribution at
energy of order $U$.

The small overshot in the sum rule is due to the fact that 
the last column is computed at the GA level without RPA
corrections, whereas
$M_0^z$ has RPA corrections to $D$. (An analogous method was
used in Ref.~\onlinecite{sei01} to compute fluctuation corrections to
$D$). The $M_0^z$ column can be considered as an improved
computation of the shielding factor. We see that the
difference is very small and for practical purposes we can use
the GA result.

From the above results it is clear that neglecting transverse
fluctuations the sum rule is well
satisfied. This is analogous to the behavior found for the spin only
model in the Ising limit (Table~\ref{tab:sw1}).


At this point we have weight at the Bragg peak and weight at energy of
order $U$. How these weights get affected by transverse fluctuations? 
On top of the covalency reduction transverse
fluctuations reduce the permanent moment (keeping the
intrinsic moment constant).
We have seen above (Sec.~\ref{sec:longitudinal-swt})
that the reduction of the moment due to transverse fluctuations
produces a concomitant large reduction of elastic spectral
weight which is transfered to multimagnon processes in the
longitudinal channel at energy $\sim J$. A direct computation in the
Hubbard model is difficult. 
In the next section we show how to estimate this effect in the Hubbard
model using the SWT results.

We expect that this low-energy rearrangement of spectral
weights will neither affect $M_{0,U}^z$ nor the shielding factor
computed above,
which, as is clear from Fig.~\ref{fig:1m2d}, is quite accurate at the
present level of approximation. Therefore we expect the present
values of $(n-2D)\mu_B^2- M_{0,U}^z$ and $M_{0,U}^z$ (without
transverse fluctuations) to be accurate estimates
of the total spectral weight at low  (order $J$)
and high (order $U$) energies.

\subsection{Effective shielding factor}
\label{sec:effect-shield-fact}

In this section we would like to show how to use the spin wave theory
results together with the results of the previous section to obtain a
better distribution of the spectral weight.

According to the discussion of  Sec~\ref{sec:1m2d},
to obtain the ``physical'' dynamical structure at low energies from
the Heisenberg model response,
one should apply the canonical transformation  back from the effective
model to the Hubbard model.
Since the processes involved in a $t/U$
expansion have a short spatial range and involve intermediate states
at high energies,  we do not expect this procedure to lead to a momentum or
low-energy dependent correction. Thus the ``physical'' dynamical correlation
functions at low energy
can be obtained from the requirement that the sum rule is
satisfied as
\begin{equation}
  \label{eq:hei2hub}
S^{\alpha\alpha}(q,\omega)=(n-2D-M_{0,U}^\alpha/\mu_B^2) S^{\alpha\alpha}_{Hei}(q,\omega)
\end{equation}
where $S^{\alpha\alpha}_{Hei}(q,\omega)$ is the dynamical correlation
function of the Heisenberg model and $M_{0,U}^\alpha$ is  the weight
transfered to high energies due to the finiteness of $U$ computed in
 Sec.~\ref{sec:hub}. Alternatively a model with higher order
 corrections (ring exchange, etc.) can be taken to
compute the response on the right hand side of Eq.~(\ref{eq:hei2hub}).

\begin{figure}[tbp]
  \centering
  \includegraphics[width=8cm,clip=true]{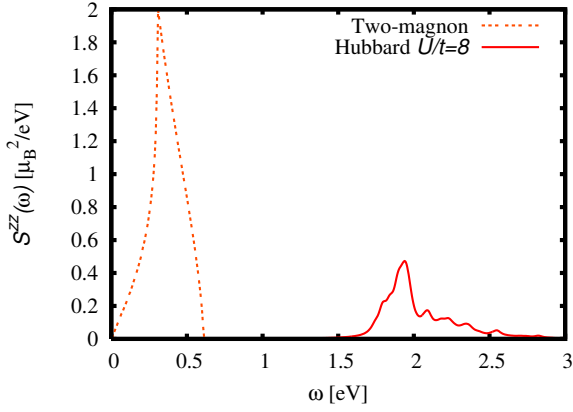}
  \caption{(Color online) Momentum integrated longitudinal spectral function
for the Hubbard model with $U=8t$, $t=335$meV and $Z_cJ=153$meV.
 We show the contribution of two-magnon excitations computed in SWT
 [Eq.~(\ref{eq:hei2hub})] with an effective shielding factor of $0.74$
(Table~\ref{tab:hub}) and $Z_{2M}=0.67$. We also show
  the contribution due to scattering across the Hubbard bands computed
 in GA+RPA.
\label{fig:2mhub}}
\end{figure}

We can consider $(n-2D-M_{0,U}^\alpha/\mu_B^2)$ in
 Eq.~(\ref{eq:hei2hub}) as an
 effective shielding factor for low-energy spectral weight.
One can check from the results of Table~\ref{tab:hub} that this
works for the Bragg intensities neglecting transverse fluctuations. Indeed we have
\begin{equation}
  \label{eq:esf}
4 m_{{\bm Q}_{AFM}}^2\mu_B^2 =(n-2D)\mu_B^2-M_{0,U}^z
\end{equation}
where the left hand side is the Bragg weight in the GA 
 for the Hubbard model without RPA correction and the right 
hand side is the effective
shielding factor times the Bragg weight in the Ising limit of the 
spin only model ($1\mu_B^2$). 

Due to the separation of energy scales
 $D$ and $M_{0,U}^z$ will be insensitive to transverse
fluctuations. The effect of the latter will be to transfer
weight from the Bragg
peak to the low-energy continua without affecting the total low-energy
spectral weight. Thus Eq.~(\ref{eq:esf}) tell us that the total low
 energy spectral weight (Bragg plus multi-magnon contribution)
is accurately given by the mean field value $4 m_{{\bm Q}_{AFM}}^2\mu_B^2$.

As a corollary we see that a good estimate of the longitudinal
effective shielding factor is given by the moment in the GA
without the need of an RPA computation. i.e. the low-energy part of
the longitudinal dynamical structure factor gets rescaled by the
mean-field value of $4 m_{{\bm Q}_{AFM}}^2$.  This
is consistent with the findings of Hirsch and Tang from numerical
data\cite{hir89} if one takes into account that the HF approximation, used by
Hirsch and Tang, gives very similar values of  $m_{{\bm Q}_{AFM}}$ as the GA
in the range of $U$ analyzed.


In Fig.~\ref{fig:2mhub} we show the two-magnon contribution computed
in the Heisenberg model and translated into a Hubbard response according to
Eq.~(\ref{eq:hei2hub}),  together with the longitudinal response at
energy $\sim U$ computed directly in the Hubbard model. 
 Additionally there is 
a Bragg peak (not shown) with weight  $16 m_{{\bm Q}_{AFM}}^2\mu_B^2
 (S-\Delta S)^2$  where $m_{{\bm Q}_{AFM}}$ is computed at mean-field
 level and  $4(S-\Delta S)^2$ takes into account the effect of
 transverse fluctuations. This is our
best estimate for the distribution of spectral weight in the
longitudinal response of the insulator. 

Notice that the
transition across the Hubbard bands appears at 2eV. One should
remember that on the mapping from the three-band Hubbard model to the
one-band Hubbard model\cite{bac91} the transition across the Hubbard
bands of the latter represent the charge transfer transitions of the
former, which according to optical  data  should occur close to
$\sim 2$eV.\cite{uch91}
Indeed in the ionic limit, the lower Hubbard band of the one-band model
corresponds to O $p^6$
states and the upper Hubbard band to Cu $d^{10}$.

\subsection{Estimation of the weights in La$_2$CuO$_4$ and
  Cu(DCOO)$_2$.4D$_2$O (CFTD)}
\label{sec:col}

In La$_2$CuO$_4$ the dynamical structure factor can be fitted\cite{col01}
in a wide range of momentum and energy by the SWT expressions
Eq.~(\ref{eq:sxaf}) with an intensity-lowering renormalization
factor  $Z_{d}^{exp}=0.51\pm 0.13$. We can use this result to estimate
the different contributions to the moments for the {\em one-magnon}
processes in La$_2$CuO$_4$:
 $$M_{NBZ}^x=0.16\pm 0.04 \mu_B^2,$$
 $$M_{MBZ}^x=0.54 \pm 0.13 \mu_B^2$$
and
 $$M_{0}^x=0.71 \pm0.18\mu_B^2.$$
The last value should be interpreted as the weight for one
transverse direction in the spin-wave like excitations of
La$_2$CuO$_4$. This is not the total weight since part of the
weight will be in higher multimagnon processes.

The experimental determination of the intensity lowering factor,
$Z_{d}^{exp}$ neglected shielding factors. Therefore the
experimentally determined quantity can be put as:
$Z_{d}^{exp}=Z_{d}^{eff}=(n-2D-M_{0,U}^x/\mu_B)Z_{d}$. Using an
effective shielding factor for $U/t=8$ as appropriate for
La$_2$CuO$_4$, $(n-2D-M_{0,U}^x/\mu_B)=0.88$ and $Z_{d}=0.57$
(SWT) one obtains $Z_{d}^{eff}=0.50$ in agreement with the
experiment (inclusion of $t'/t=-0.2$ does not change
this number appreciably). 

Subtracting the observed one-magnon
weight from the total expected transverse weight of $0.88\mu_B $
would then leave a total of $0.17 \pm 0.18\mu_B^2$ in three, five and
higher multi-magnon processes which is within the error bars as 
illustrated in Fig.~\ref{fig:n2ddx}. 
Notice that the detected one-magnon spectral
weight is significantly lower than $1\mu_B$, the value expected
from the sum-rule neglecting the shielding factors and quantum
corrections.

For Cu(DCOO)$_2$.4D$_2$O (CFTD) R{\o}nnow {\it et al.}\cite{ron01}
also find $ Z_{d}^{exp}= 0.51\pm 0.04$ within LSWT. For this
compound the ratio of $U/t$ is much larger than in the cuprates
and the shielding factor correction should therefore be close to
1. Since the error bars are smaller than in the cuprates this
experiment shows, not surprisingly, that multimagnon processes are
needed to satisfy the sum rule (see Fig.~\ref{fig:n2ddx}). The
similarity for $Z_{d}^{exp}$ found for the two compounds is to
some extent unexpected since there should be a difference due to
the shielding factors, however, the expected difference is within
the current experimental errors. Notice also that $g$ may also
differ in the two compounds.

\section{Away from half-filling}
\label{ahf}
\subsection{Shielding factors}
Taking $g=2$ and defining the doping as $x=1-n$,
the shielding factor for general doping changes from $(1-x^2)/2$
for $U=0$ to $ 1-|x| $ for $U=\infty$, the latter
form being used often in experimental works.
In general we expect a dependence on $U/t$
qualitatively similar to the one shown at $x=0$ (c.f. Fig.~\ref{fig:1m2d}).

In order to proceed in the doped phase
we need a model for the ground state. In this work we are interested in
overall distributions of weights and we expect this to be
to a large extent insensitive to the details of the ground state if
correlations are taken reasonably well into account.
This can be seen already at $x=0$ (c.f. Fig.~\ref{fig:1m2d}) where we
see that the paramagnet and the SDW within the GA give similar shielding
factors although the nature of the ground state is completely
different.
Detailed distributions of spectral weight as measured in
Ref.~\onlinecite{tra04} and computed in
Ref.~\onlinecite{sei05} will, of course, depend
strongly on the ground state and so can be quite helpful
to determine it. Since those issues are beyond our present scope
 we restrict the description of the mean-field states to a minimum.

For the specific system La$_{2-x}$X$_x$CuO$_4$ (X=Sr,Ba)
 we use a ground state
consisting of stripes as suggested by experiment.\cite{tra95,tra04}
Metallic stripes parallel to the CuO bond can be obtained within the GA for
the Hubbard model where a next-nearest neighbor hopping $t'/t=-0.2$ has to
be implemented in order to have one doped hole per every second unit
cell along the stripe \cite{sei04a} (so called
``half-filled'' stripes) in agreement with experiments.
Note that a similar value for $t'/t$  is predicted by first principle
computations \cite{pav01} for La$_{2-x}$Sr$_x$CuO$_4$.

It should be mentioned that the results at half-filling
reported in the previous sections are rather insensitive to
the next-nearest neighbor hopping.
For $U/t=8$ and $t'/t=-0.2$ we obtain an analogous dispersion than that
reported in Fig. \ref{fig:edq} with the only difference that
the value for the nearest neighbor hopping has to be scaled
to $t=353.7$meV. This latter parameter set has been
used\cite{sei05} in order to explain a recent MNS experiment
on Ba-codoped LSCO.\cite{tra04} Overall  spectral weights with this extended
parameter set are practically the same than those reported in
Sec.~\ref{sec:hubtrans} with $t'=0$.

In Fig.~\ref{fig:n2ddx} we show the shielding factor obtained 
averaging ($n-2D$) in the mean-field solutions
as a function of doping.
 We see that the shielding factor has a linear behavior
similar to the one for the $U/t=\infty$ case. As expected we find
that the shielding factor is similar for other low-energy textures
(not shown). For doping $x=1/8$ we obtain $n-2D=0.79$
so there is an extra 10\% reduction with respect to the undoped case.

\subsection{Spectral weights}

\subsubsection{Longitudinal Part}

Because the stripe solutions are magnetic 
the longitudinal structure factor has a Bragg  part and an
inelastic part [c.f. Eq.~(\ref{eq:szafh})].

Since the stripe
solutions are metallic, low-energy particle-hole excitations
are allowed. Therefore contrary to the result
in the insulator (Sec.~\ref{sec:hfhublon}),
 already at RPA level one finds weight at magnetic
energies ($\lesssim 0.3$eV) as shown in Fig.~\ref{fig:szsxsy}.
This weight however is broadly distributed in momentum space and
will most likely pass unnoticed in unpolarized INS as discussed
in Sec.~\ref{sec:exper}.
Just as shown in Fig.~\ref{fig:2mhub} for
the insulator a more elaborate computation, taking into account
longitudinal fluctuations, will show in addition  weight
at the two-magnon excitations. Since the inelastic longitudinal
component due to multimagnon scattering is already quite featureless  in the
insulator (Sec.~\ref{sec:longitudinal-swt} and \ref{sec:hfhublon})
it is reasonably  to expect that it will be even more
featureless in the doped phase. We conclude that within standard
protocols (Sec.~\ref{sec:exper}) to first approximation all the inelastic
longitudinal spectral weight will be assigned to the background.

\begin{figure}[tp]
  \centering
   \includegraphics[width=8cm,clip=true]{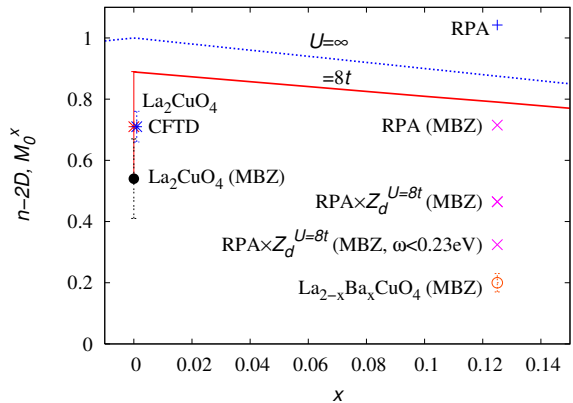}
\caption{(Color online)
Zeroth moment and shielding factor as a function of
doping. We show the shielding factor  for  $U/t=8$ and
$t'/t=-0.2$ for the SDW at $x=0$ and bond-centered metallic stripes solutions
as a function of doping (red line)
 and for  $U/t=\infty$ (dotted blue line). For the zeroth moment
MBZ means that the integration was restricted to the magnetic
Brillouin zone.  La$_2$CuO$_4$\cite{col01} and CFTD\cite{ron01} label
the moments estimated in
insulators (Sec.~\ref{sec:col}) and
La$_{2-x}$Ba$_x$CuO$_4$\cite{tra04} labels
the experimental value of $M_0$ in the magnetic Brillouin zone,
corrected by the
polarization factor. In the last case the error bars reflect the
two extreme possibilities for the polarization factor 2/3 and 1/2
(c.f. Sec.~\ref{mns}).  The relative error should be
larger than at zero doping where the error bars have the usual sense.
Finally we show the zeroth moment of the GA+RPA transverse
spectra at $x=1/8$ in the whole Brillouin zone (+) and in the
MBZ $\times$. The two lower  $\times$ include a
$Z_{d}^{U=8t}=0.65$ renormalization factor and for the lower one the
energy integration was restricted to energies smaller than 0.23eV.
 \label{fig:n2ddx}}
\end{figure}

\begin{figure}[htbp]
  \centering
      \includegraphics[width=8cm,clip=true]{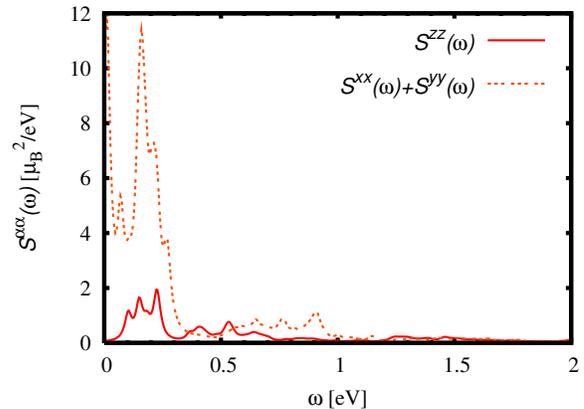}
  \caption{(Color online) Longitudinal and transverse components of the spin
      autocorrelation function for $x=1/8$ in a $16\times 4$ site
      cluster for $U/t=8$ and $t'/t=-0.2$ and $t=353.7$meV. Only the
      inelastic  part is shown and we neglect the intensity
      renormalization ($Z=1$). The ground state consist of 4 SC
      stripe running along the short
      dimension.
 }
  \label{fig:szsxsy}
\end{figure}

For the specific case of a striped ground state we can also estimate the
elastic spectral weight in the longitudinal channel. For doping $x=1/8$
and an array of $d=4$ SC stripes we find elastic peaks at
${\bm Q}_1=(1/2 \pm \epsilon ,1/2)2\pi$ and ${\bm Q}_2=(1/2 \pm 2 \epsilon
,1/2)2\pi$ with $m_{{\bm Q}_1}=0.215$ and very small weight, $m_{\bm Q}^2$,
in the higher harmonics $m_{{\bm Q}_2}=0.018$.

It was shown in the insulator that if one neglects transverse
fluctuations the sum rule in the longitudinal channel is exhausted 
with a good degree of approximation by the Bragg weight plus the inelastic high
energy  weight. 
Is there an analogous behavior  in the doped phase?  The elastic weight
neglecting transverse fluctuations is  $0.37\mu_B^2$ and the
 inelastic longitudinal spectral weight is $0.59 \mu_B^2$. This
 includes low-energy particle-hole excitations and weight due to
 transitions across the Hubbard bands. The
total weight in the longitudinal channel is $0.96  \mu_B^2$ to be
compared with a shielding factor of  0.79.  In this case we find an
overstrike of the sum rule which is more sever than in the insulator
and which is even larger than the maximum allowed value $1-x=0.875$.
This is because RPA becomes less accurate in the metallic phase due to
the presence of small energy denominators. The violation however is
not as severe as for the transverse channels as reported below.

The elastic weight neglecting transverse fluctuations is
roughly half of the one found in the insulator ($0.74 \mu_B$).
Just as in the insulator a large
fraction of this weight (at least 60\%) will be transfered to multimagnon
excitations if transverse fluctuations are taken into account. It is
conceivable that all the weight in the Bragg peak is transfered in which
case the system becomes quantum critical or quantum disordered.\cite{sac01}

\subsubsection{Transverse Part}

The transverse component of the dynamical structure factor
shows sharp features due to the propagating spin-wave like modes of the
stripes. In addition, since the stripes are metallic, a fraction of the
spectral weight is in a broad particle hole continuum. The latter
features however are mainly located in the nuclear
Brillouin zone (NBZ).\cite{sei05}

In the magnetic Brillouin zone (MBZ) we find mainly
sharp propagating collective modes which are good candidates to be
easily detected. These collective modes are analogous to the
spin-wave  modes of the insulator and therefore we expect that their
spectral weight will be overestimated in RPA and should be corrected
by a quantum renormalization $Z_d^U$ as in the insulator.
 $Z_d^U$ may be smaller than in the insulator due to stronger
quantum fluctuations. On the other hand for a heavily doped system
one should recover the noninteracting value $Z_d=1$. Since doping is
small we tentatively take the same value as in the insulator $Z_d^{U=8}=0.65$.

\begin{figure}[tbp]
      \includegraphics[width=8cm,clip=true]{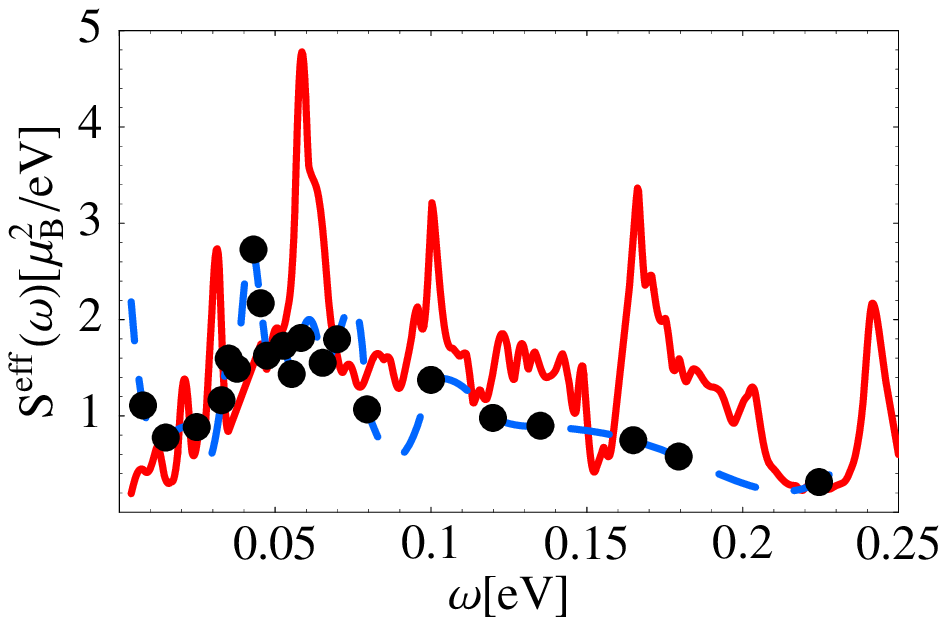}
      \includegraphics[width=8cm,clip=true]{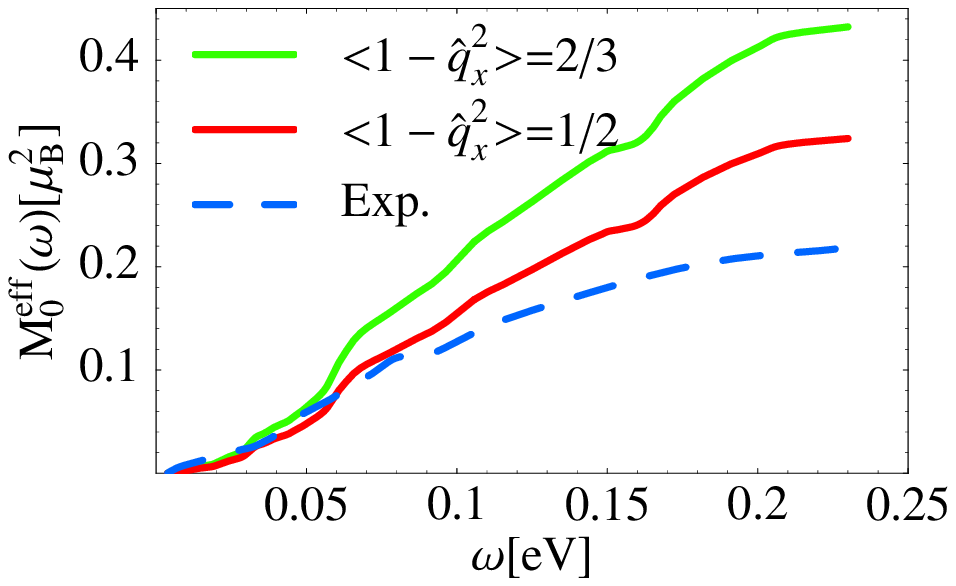}
  \caption{(Color online) The upper panel shows the theoretical and the experimental
result for $S^{eff}(\omega)$ using $U/t=8$, $t'/t=-0.2$, $t=353.7$eV and
$(1-\hat q_x^2)\rangle_{dom} =(1-\hat q_y^2)\rangle_{dom} =1/2$,
in a $96\times96$ sites lattice,
together with the experimental data for La$_{2-x}$Ba$_x$CuO$_4$
$x=0.125$\cite{tra04}. The lower panel shows the partial sum rule
weight.
$M_0^{eff}(\omega)\equiv\frac{1}{N} \sum_{\bm{q}}\int_{-0}^{\omega}
d(\hbar\omega') S^{eff} (\bm{q},\omega')$
}
  \label{fig:seffexptheo}
\end{figure}

The momentum distribution of the transverse spectral weight has been shown in
Ref.~\onlinecite{sei05}.
In Fig.~\ref{fig:n2ddx} we show the transverse integrated spectral weight
compared with the experimental results. At zero doping the spectral
weight coincides with the theoretical value (not shown for clarity)
as discussed in Sec.~\ref{hf}. At finite doping the weight in the
MBZ decreases roughly as  the shielding factor as can be seen by
comparing the points labeled La$_2$CuO$_4$(MBZ) and RPA$\times
Z_d^{U=8t}$(MBZ). Comparing Fig.~\ref{fig:swt} and Fig.~\ref{fig:szsxsy} we
see that whereas in the insulator almost all of the spectral weight is at
magnetic energies in the doped phase a large fraction of the spectral
weight is at intermediate energies not accessibly to MNS. In other
words doping induces a transfer of spectral weight from low energies
to intermediate  energies in this channel. Interestingly  optical
spectra shows that doping also generates structure at the same
energies in the charge channel\cite{uch91,lor93a,lor03}.

The lower ``$\times$'' at $x=1/8$ in Fig.~\ref{fig:n2ddx} is the
RPA moment in the MBZ renormalized by $Z_{d}^{U=8t}$ and restricting
the integral only up to the highest energy measured in
Ref.~\onlinecite{tra04}, $\omega <0.23$eV. We see that the obtained
value is roughly 50\% higher than the experimental weight.

%

In order to see at what energy the disagreement arises
we show in Fig.~\ref{fig:seffexptheo} the experimental $S^{eff}$
compared with the theoretical one using 1/2 as polarization factors
[Eq.~(\ref{eq:seffqz})] and neglecting the longitudinal weight.
The lower panel shows a comparison of the partial sum rule weight with
the two limiting values of the polarization factor. The theoretical
computation shows a strong resonance at $\omega=58$meV
which in the experiment  appears with a reduced spectral weight and in part
distributed on a broader energy range. (The resonance appears at
higher energy if one looks only at the $Q_{AFM}$
response\cite{sei05}).
The overall weight integrated up to energies immediately above the
resonance is in agreement with the experimental one. As the energy
increases there are strong deviations between theory and experiment.

The simplest explanation for the smaller weight in the experiment is
that some of the structures that contribute at high energy may be
broad and assigned to the background (see Ref.~\onlinecite{sei05}).
We should say that because we neglect life-time effects even the sharp
structures of our computation will be  much broader in reality
and this will be more important as the energy increases since, quite
generally, the phase space for decay processes increases.
 Also high-energy excitations will be much more sensitive
to disorder since they involve short wave-lengths and this will also
tend to make them broader. The broad structures will be assigned to
the background explaining the difference between theory
and experiment.
We should also take into account that $Z_d$ can be more depressed
due to the larger impact of quantum fluctuations in the metallic
phase.

\begin{figure}[t]
  \centering
      \includegraphics[width=8cm,clip=true]{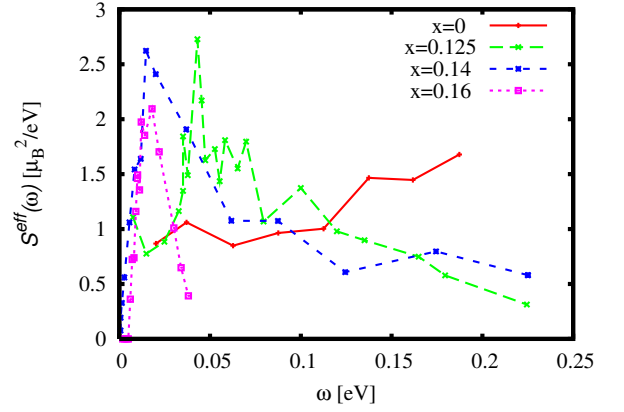}
  \caption{(Color online) Experimental results for $S^{eff}(\omega)$
for
$x=0$     (Ref.~\onlinecite{hay96}),   La$_{2-x}$Ba$_x$CuO$_4$
$x=0.125$ (Ref.~\onlinecite{tra04}) and  La$_{2-x}$Sr$_x$CuO$_4$ with
$x=0.14$  (Ref.~\onlinecite{hay96})
and
$x=0.16$ (Ref.~\onlinecite{chr04}). We used the following conversions
from the quantity reported in the quoted reference:
$S^{eff}=S$ (Ref.~\onlinecite{tra04})
$=\frac{2}\pi\chi $(Ref.~\onlinecite{hay96})
$=\frac{2}3\chi $ (Ref.~\onlinecite{chr04}).}
  \label{fig:seffexp}
\end{figure}

In order to have a broader view of the trends in the evolution of the spectral
weight, we show  in  Fig.~\ref{fig:seffexp} a compilation of
measurements of $S^{eff}$
from different groups and at different dopings.
For $x=0, 0.125$ and
0.14 the overall weight seem to be roughly conserved whereas for $x=0.16$  the
detected weight seems to be significantly smaller. It will be interesting
to see  weather this is due to a transfer to high-energies where absolute
measurements are not available.


\section{Conclusions}
\label{con}

In this work we have analyzed the distribution of spectral weight
in neutron scattering experiments in cuprates.

In the insulator
the spectral weight has been estimated in detail in the Heisenberg
model and then translated into a Hubbard response with the use of an
effective shielding factor. We find an effective shielding factor
which essentially coincides with a proposal of Hirsch and Tang
obtained by analyzing numerical data.\cite{hir89} In addition
we have estimated, with the aid of the TDGA, the weight at the
energy of the charge transfer  transitions.

The inelastic longitudinal spectral weight at low energies is in
broad features and thus hard to detect, although progress has recently
been made.\cite{hub05}
The transverse part has the well know quantum renormalized 
propagating spin wave modes
which will be the features more accessible to experiment. Theory
and experiment agree well in the insulator.

For the doped phase we computed the dynamical response within the TDGA.
We considered metallic stripes but we expect our results for overall
weights to be largely insensitive to the specific texture.
One finds extra broad features due to the particle-hole continua plus
propagating collective modes. In analogy with the insulator we expect
the latter to be reduced in intensity due to quantum fluctuation and to
dominate the spectral weight reported in experiments.

Our motivation was to
understand why such a small fraction of the naively expected spectral
weight is actually detected.
We see that after taking into account all
the reduction factors we passed from a weight which was an order of
magnitude smaller than expected to a weight which
is $\sim2/3$ of that expected. Given the uncertainties involved we
think this value is reasonable.

By far the most important factor in reducing the detectable weight turns out
to be the fact that a large fraction of the spectral weight is in
broad features or at too high energies and this make the detection of
the total spectral weight experimentally very challenging.
In the doped phase a consistent fraction of the transverse spectral weight
is predicted to lay at intermediate energies which may become
accessible to experiment in a near future.

We see no reason why the broad/high energy modes can not
have a strong impact on the effective attraction between holes relevant for
superconductivity and therefore their experimental characterization remains
an important open problem.

\acknowledgments
We acknowledges support from the UK EPSRC under grant
number GR/R76714/01 (RC). JL thanks the University of Oxford,
ISIS and the CCLRC's Rutherford Appleton Laboratory for financial
support and hospitality during part of this work and  A. T. Boothroyd,
S. M. Hayden, D.F. McMorrow, T. G. Perring and J. Tranquanda 
for illuminating discussions.


\begin{thebibliography}{10}

\bibitem{tra04}
J.~M. Tranquada {\it et~al.}, Nature (London) {\bf 429},  534  (2004).

\bibitem{sca98}
D.~J. Scalapino and S.~R. White, Phys.\ Rev.\ B {\bf 58},  8222  (1998).

\bibitem{dem98}
E. Demler and S.-C. Zhang, Science {\bf 396},  733  (1998).

\bibitem{dai99}
P. Dai {\it et~al.}, Science {\bf 284},  1334  (1999).

\bibitem{dai00}
P. Dai {\it et~al.}, Nature (London) {\bf 406},  965  (2000).

\bibitem{kee02}
H.-Y. Kee, S.~A. Kivelson, and G. Aeppli, Phys.\ Rev.\ Lett. {\bf 88},  257002
  (2002).

\bibitem{sei01}
G. Seibold and J. Lorenzana, Phys.\ Rev.\ Lett. {\bf 86},  2605  (2001).

\bibitem{sei03}
G. Seibold, F. Becca, and J. Lorenzana, Phys.\ Rev.\ B {\bf 67},  085108
  (2003).

\bibitem{sei04b}
G. Seibold, F. Becca, P. Rubin, and J. Lorenzana, Phys.\ Rev.\ B {\bf 69},
  155113  (2004).

\bibitem{lov84}
S.~W. Lovesey, {\em Theory of Neutron Scattering form Condensed Matter}
  (Clarendon Press, Oxford, 1984).

\bibitem{wal90}
R.~E. Walstedt and W.~W. {Warren, Jr.}, sci {\bf 248},  1082  (1990).

\bibitem{can92p10131}
C.~M. Canali, S.~M. Girvin, and M. Wallin, Phys.\ Rev.\ B {\bf 45},  R10131
  (1992).

\bibitem{qmc}
We neglect the small $~7\%$ variation of $Z_C(\bm{q})$ along the
  antiferromagnetic zone boundary predicted by Ref. \onlinecite{can92p10131}
  and R.R.P. Singh and M. Gelfand, Phys. Rev. B {\bf 52}, R15695 (1995); O.F.
  Sylju\aa{}sen and H.M. R\o{}nnow, J. Phys.: Condens. Matter {\bf 12}, L405
  (2000).

\bibitem{sin89p9760}
R.~R.~P. Singh, Phys.\ Rev.\ B {\bf 39},  R9760  (1989).

\bibitem{man91}
E. Manousakis, Rev.\ Mod.\ Phys. {\bf 63},  1  (1991).

\bibitem{hei81}
I. Heilmann {\it et~al.}, Phys.\ Rev.\ B {\bf 24},  3939  (1981).

\bibitem{hub05}
T. Huberman {\it et~al.}, phys. Rev. B (2005, in press) (unpublished).

\bibitem{can93}
C.~M. Canali and M. Wallin, Phys.\ Rev.\ B {\bf 48},  3264  (1993).

\bibitem{san01}
A.~W. Sandvik and R.~R.~P. Singh, Phys.\ Rev.\ Lett. {\bf 86},  528  (2001).

\bibitem{per93}
J.~D. Perkins {\it et~al.}, Phys.\ Rev.\ Lett. {\bf 71},  1621  (1993).

\bibitem{lor95}
J. Lorenzana and G.~A. Sawatzky, Phys.\ Rev.\ Lett. {\bf 74},  1867  (1995).

\bibitem{lor95b}
J. Lorenzana and G.~A. Sawatzky, Phys.\ Rev.\ B {\bf 52},  9576  (1995).

\bibitem{kas98}
M.~A. Kastner, R.~J. Birgeneau, G. Shirane, and Y. Endoh, Rev.\ Mod.\ Phys.
  {\bf 70},  897  (1998).

\bibitem{gru00}
M. Gr{\"u}ninger {\it et~al.}, Phys.\ Rev.\ B {\bf 62},  12422  (2000).

\bibitem{lor99}
J. Lorenzana, J. Eroles, and S. Sorella, Phys.\ Rev.\ Lett. {\bf 83},  5122
  (1999).

\bibitem{can92p7127}
C.~M. Canali and S.~M. Girvin, Phys.\ Rev.\ B {\bf 45},  7127  (1992).

\bibitem{fan90}
G. Fano, F. Ortolani, and A. Parola, Phys.\ Rev.\ B {\bf 42},  R6877  (1990).

\bibitem{vol84}
D. Vollhardt, Rev.\ Mod.\ Phys. {\bf 56},  99  (1984).

\bibitem{yos96}
K. Yosida, {\em Theory of Magnetism} (Springer-Verlag, Berlin, 1996).

\bibitem{col01}
R. Coldea {\it et~al.}, Phys.\ Rev.\ Lett. {\bf 86},  5377  (2001).

\bibitem{sei05}
G. Seibold and J. Lorenzana, Phys.\ Rev.\ Lett. {\bf 94},  107006  (2005).

\bibitem{rog89}
M. Roger and J.~M. Delrieu, Phys.\ Rev.\ B {\bf 39},  2299  (1989).

\bibitem{sch90}
H.~J. Schmidt and Y. Kuramoto, Physica C {\bf 167},  263  (1990).

\bibitem{lem97}
F. Lema, J. Eroles, C.~D. Batista, and E.~R. Gagliano, Phys.\ Rev.\ B {\bf 55},
   15295  (1997).

\bibitem{sei04a}
G. Seibold and J. Lorenzana, Phys.\ Rev.\ B {\bf 69},  134513  (2004).

\bibitem{pav01}
E. Pavarini,  I.  Dasgupta,  T. Saha-Dasgupta, O. Jepsen and O.K. Andersen, Phys.\ Rev.\ Lett. {\bf 87},  047003  (2001).

\bibitem{hir89}
J.~E. Hirsch and S. Tang, Phys.\ Rev.\ Lett. {\bf 62},  591  (1989).

\bibitem{bac91}
S. B. Bacci, E.~R. Gagliano, R~M. Martin, and J.~F. Annett, Phys.\ Rev.\ B {\bf 44},  7504 (1991).

\bibitem{uch91}
S. Uchida {\it et~al.}, Phys.\ Rev.\ B {\bf 43},  7942  (1991).

\bibitem{ron01}
{H.M. R{\o}nnow {\it et al.}}, Phys.\ Rev.\ Lett. {\bf 87},  037202  (2001).

\bibitem{tra95}
J.~M. Tranquada {\it et~al.}, Nature (London) {\bf 375},  561  (1995).

\bibitem{sac01}
S. Sachdev, {\em Quantum phase transitions} (Cambridge University Press,
  Cambridge, 2001).

\bibitem{lor93a}
J. Lorenzana and L. Yu, Phys.\ Rev.\ Lett. {\bf 70},  861  (1993).

\bibitem{lor03}
J. Lorenzana and G. Seibold, Phys.\ Rev.\ Lett. {\bf 90},  066404  (2003).

\bibitem{hay96}
S.~M. {Hayden {\it et al. }}, Phys.\ Rev.\ Lett. {\bf 76},  1344  (1996).

\bibitem{chr04}
N.~B. Christensen {\it et~al.}, Phys.\ Rev.\ Lett. {\bf 93},  147002  (2004).

\end{thebibliography}

 \end{document}